\documentclass[12pt]{article}
\usepackage[english]{babel}
\usepackage[pdftex]{color}
\usepackage{amsmath}
\usepackage{graphicx}
\usepackage{hyperref}
\usepackage{amsmath}
\usepackage{amsfonts}
\usepackage{amssymb}

\begin{document}
\date{}

\begin{center}
{\Large\textbf{{}BRST-BV approach for interacting higher
spin fields}} \vspace{18mm}

{\large  A.A. Reshetnyak$^{(a,b,c)}\footnote{E-mail:
reshet@tspu.edu.ru}$}

\vspace{3mm}

\noindent  ${{}^{(a)}} ${\em
Center of Theoretical Physics, \\
Tomsk State Pedagogical University,\\
634061 Tomsk, Russia}

\noindent  ${{}^{(b)}} ${\em
National Research Tomsk State  University,\\
634050 Tomsk, Russia}

\noindent  ${{}^{(c)}} ${\em
National Research Tomsk Polytechnic   University,\\
634050 Tomsk, Russia} \vspace{10mm}

\begin{abstract}
We develop the BRST-BV approach to construct the general off-shell
 Lorentz covariant cubic, quartic,  $e$-tic interaction vertices for irreducible  higher spin fields on $d$-dimensional Minkowski space. We
consider two different cases for interacting  integer  higher spin fields both with massless and with massive fields. The deformation procedure to find  minimal (determined with help of generalized Hilbert space)  BRST-BV  action   for interacting higher spin fields is based on the preservation of master equation validity  in each power of coupling constant $g$ starting from the Lagrangian formulation for free   gauge theory.
   As examples we consider the construction of local cubic vertices for $k$ irreducible  massless fields   of integer
helicities, and $k-1$   massless with one massive fields of spins $s_1, ,..., s_{k-1}, s_k$.  For triple of two  massless scalars and tensor field of integer spin the BRST-BV action with  cubic interaction is explicitly found.  Unlike the previous results on cubic vertices
we follow our result  for BRST approach \cite{BRcub} for massless fields, but for the unique  BRST-BV action instead of classical action with reducible gauge transformations.  The procedure is based on  the complete BRST operator, including the trace constraints  that is
used to formulate an irreducible representation with definite
integer spin.  \end{abstract}

\end{center}
\thispagestyle{empty}

\section{Introduction}

The assumption that the higher spin field  theory can open possibilities for  new physics beyond the Standard Model and contribute to the formulation of quantum gravity due to matter particles and interaction carriers  with higher spins is one of the most attractive in modern high-energy theoretical  physics for constructing interactions. The rapt attention for higher spin fields is because of its close relation with (Super)string Field Theory (see for a review, e.g. \cite{revvas},  \cite{revBCIV},
\cite{reviews3}, \cite{rev_Bekaert}, \cite{reviewsV},
\cite{Snowmass}, \cite{Ponomarev} and the references therein).

The  structure of cubic and quartic vertices for different  higher
 spin fields  have been investigated by {many authors} with use of different methods
(see, e.g., the recent papers with the references therein for the cubic \cite{Manvelyan}, \cite{Manvelyan1},
\cite{Joung}, \cite{frame-like1},  \cite{Tsulaiai2009}
\cite{BRST-BV3},  \cite{frame-like2}, \cite{BKTW},
\cite{Metsaev-mass}, \cite{BRcub}, \cite{Rcubmasless}, \cite{BRcubmass},  \cite{BKStwis}, \cite{SkvortsovTungMiriam} and for quartic \cite{T1}, \cite{DT},  \cite{L1}).
We stress, the results on the structure of cubic vertices
obtained  in terms of physical degrees of freedom in a concise form
in the light-cone {formalism} in \cite{Metsaev0512},
\cite{Metsaev-mass}. In the covariant metric-like form the list of
cubic vertices for reducible representations of Poincare group with
discrete spins (being consistent with \cite{Metsaev0512}) are
contained in \cite{BRST-BV3}, where the cubic vertices were derived
using the constrained  BRST approach, but without imposing on the
vertex  the algebraic  constraints.
%The latter peculiarity
 %leads to the violation of the irreducibility of the
%representation for interacting higher spin fields and, hence, to an
%undesirable change of the number of physical degrees of freedom.
Also, we point out the constructions of cubic vertices
within the BRST approach  without use of constraints responsible for
trace conditions in the BRST operator. i.e. for reducible higher spin fields  (see e.g. \cite{BKTW} and the
references therein).

At the same time the formulation of the BRST-BV approach to construct minimal BV (Batalin-Vilkovisky)  action \cite{BV}, \cite{BV1}, \cite{BV-GA} which encodes gauge algebra for the gauge model with (ir)reducible higher spin fields on constant curvature spaces in question with complete BRST operator has not been yet suggested, in spite of the results for constrained BRST-BV approach for the fields with higher integer spin: \cite{BGST}, \cite{AGT}, \cite{BRST-BV3}; with half-integer spins \cite{conHIBRST-BV} and with integer continuous spin \cite{BPR}. The construction of BRST-BV minimal actions may be used both for  deformation procedure of finding the interacting vertices within string-inspired oscillator formalism and to construct quantum gauge-fixed BRST-BV action and path integral, firstly, suggested for totally symmetric constrained higher integer spin fields on the flat space-times in \cite{2010.15741}.

    In this paper, we formulate BRST-BV approach for irreducible massless and massive higher integer spin fields in $d$-dimensional Minkowski space with complete BRST operator, suggest the deformation procedure to find deformed gauge algebra for $k$-samples of   free higher integer spin fields, which includes the interacting vertices for the action (starting from cubic), generators and structure functions for the   gauge transformations (starting from linear and zeroth approximations in fields).
           Then we explicitly  derive the cubic vertices for irreducible massless and massive
higher spin fields focusing on the manifest Poincare covariance.
Research is realized with partial using the BRST approach with complete BRST
operator, which (following the method \cite{BRcub},
\cite{Rcubmasless},  \cite{BRcubmass} see as well the results for the massless fields
with two-component spinor indices in $4d$ flat space
\cite{BKStwis})  contains converted set of operator constraints
forming a first-class gauge algebra. The set of constraints includes
on equal-footing the on-shell condition $l_0$ and constraints $l_1,
\, l_{11}$, responsible for divergences and traces. The
operator $l_{11}$ is consistently imposed when using the constrained
BRST (and BRST-BV) approach on the set of fields and gauge parameters as the
holonomic constraints for simplicity of calculations  beyond  the
Lagrangian formulation. This approach inherits the way of obtaining
the Lagrangian formulation for higher spin fields  from the
tensionless limit \cite{BRST-BFV1} for (super)string theory with
resulting BRST charge without presence of the algebraic (e.g. trace)
constraints. We are again  repeating following to  \cite{BRcub}, \cite{BRcubmass} that this procedure is
correct but the actual Lagrangian description of irreducible fields
is realized only after {additional} imposing the subsidiary
conditions which are not derived from the Lagrangian. Of course, the
Lagrangian formulations for the same irreducible field with higher
(half)-integer spin in Minkowski space obtained in constrained BRST
approach and  BRST approach with complete BRST operator describe the
same  particle on free level due to equivalence between these
formulations \cite{Reshetnyak_con}. This equivalence, automatically, follows for the minimal BRST-BV actions
 for the same irreducible field  obtained from BRST-BV approach with complete BRST operator and one with constrained BRST-BV approach.   However, the same equivalences have not
yet been established for interacting irreducible  higher-spin fields
as it was recently demonstrated for cubic vertices in massless \cite{BRcub},
\cite{Rcubmasless} and in massive cases \cite{BRcubmass}.

As a result, we again face the problem  when constructing the
 general  covariant vertices  but now within BRST-BV approach
with complete BRST operator  for irreducible  higher
integer spin fields on $d$-dimensional flat space-time within
metric-like formalism. It is exactly one from the
problems that it  will be solved in the paper.

Because  the minimal BRST-BV action (satisfying master equation) encodes the gauge algebra with gauge functions
 from the gauge-invariant  Lagrangian formulation obtained  within BRST approach   for the same interacting
higher integer spin fields, the structure of the vertices should have similar form, e.g. as for cubic vertices in   \cite{BRcub},
\cite{Rcubmasless}, \cite{BRcubmass}.
 Thus, again, we  expect the
resulting cubic vertices will contain new terms (as compared with
\cite{BRST-BV3}) with the trace constraints  and may have various
representations.

Note, the BRST approach with complete BRST operator to a
Lagrangian description of various free and interacting
higher spin field models in Minkowski and AdS spaces has been
developed in many works (e.g., see the papers  \cite{BPT},
\cite{BKP}, \cite{BKr}, \cite{BFPT}, \cite{BGK},  \cite{BKR},
\cite{BR}, and the review \cite{reviews3}). The aim of the paper is
to develop BRST-BV approach with complete BRST operator;  to formulate the
deformation procedure to determine cubic, quartic and so on  vertices for       interacting
higher spin field; to
present a complete solution of problem for the cubic vertices for
unconstrained massless and massive higher spin fields and to obtain from general oscillator-like vertices
explicit tensor representation for BRST-BV minimal action for some  triple of massless interacting higher
spin fields.

The paper has the following organization.  Section~\ref{cBRSTBFV}
presents the basics of a  BRST-BV approach for minimal BRST-BV action  construction for free
 totally symmetric   higher spin field, with all constraints $l_0,\, l_1,\,
l_{11}$ taken into account. In Section~\ref{BRSTinter}, we formulate deformation procedure for minimal BRST-BV action   in power of coupling constant $g$ based  on the master equation validity.
Then, we deduce a
system of equations for a $e$-tic, $e=3,4,...$  deformation in fields and antifields  of
quadratic  free BRST-BV action.  A solution for the
deformed cubic vertices and gauge transformation incorporated in cubic approximation for minimal BRST-BV action  is given in a
Section~\ref{BRSTsolgen} for $k$  massless,  one massive and $(k-1)$  massless fields. The  example for the massless fields with  special
set of spins is presented in the Section~\ref{examples}.   In conclusion  a
final summary with comments are given.

 The usual definitions and notations  from the work
\cite{BRcub} are used  for a metric tensor $ \eta_{\mu\nu} = diag (+,
-,...,-)$ with Lorentz indices $\mu, \nu = 0,1,...,d-1$ and the
respective notation $\epsilon(F)$, $(gh_H,gh_L, gh_{\mathrm{tot}})(F)$, $[F,\,G\}$, $[x]$,
$(s)_{k}$, $\theta_{m,0}$  for the values of Grassmann parity and Hamiltonian, Lagrangian, total $gh_H+gh_L=gh_{\mathrm{tot}}$   ghost numbers of a
homogeneous quantity $F$, as well as the supercommutator,  the
integer part of a real-valued $x$, for  the  integer-\-valued vector  $
(s_1,s_2,...,s_k)$ and Heaviside $\theta$-symbol ($\theta_{m,0}=1(0)$ for $m>0$ $(0\geq m)$).

\section{BRST-BV approach for  free field with integer  spin}

\label{cBRSTBFV}%%%%%%%%%%%%%%%%%%%%%%%%%%%%%%%%%%%%%%%%%%%%%%%%%%%%%%%%%%%%%%%%%

Here, we present the basics  of the BRST approach and develop the BRST-BV approach  to free
massless and massive higher integer spin field theory for its following use to
construct a general $n$-tic  interacting vertices initializing from the BRST approach.

The unitary massless  (massive)  Poincare  group  irreducible representations
with integer helicities (spins) $s$  can be
realized using the real-valued totally symmetric tensor fields
$\phi_{\mu_1...\mu_s}(x)\equiv \phi_{\mu(s)}$ subject to the
conditions
\begin{eqnarray}\label{irrepint}
    &&  \big(\partial^\nu\partial_\nu + \theta_{m,0}m^2,\, \partial^{\mu_1},\, \eta^{\mu_1\mu_2}\big)\phi_{\mu(s)}  = (0,0,0)  \  \ \ \   \Longleftrightarrow  \  \\
     &&       \big(l_0,\, l_1,\, l_{11}, g_0 -d/2\big)|\phi\rangle  = (0,0,0,s)|\phi\rangle. \nonumber
\end{eqnarray}

The basic vector $|\phi\rangle$ and the operators $l_0,\,
l_1,\, l_{11}, g_0$ above  are defined in the Fock space $\mathcal{H}$
with the Grassmann-even oscillators $a_\mu, a^+_\nu$, ($[a_\mu, a^+_\nu]= - \eta_{\mu\nu}$)  as follows
\begin{eqnarray}\label{FVoper}
&&   |\phi\rangle  =  \sum_{s\geq 0}\frac{\imath^s}{s!}\phi^{\mu(s)}\prod_{i=1}^s a^+_{\mu_i}|0\rangle, \\
&&   \big(l_0,\, l_1,\, l_{11}, g_0\big) = \big(\partial^\nu\partial_\nu+ \theta_{m,0} m^2 ,\, - \imath a^\nu  \partial_\nu ,\, \frac{1}{2}a^\mu a_\mu ,  -\frac{1}{2}\big\{a^+_{\mu},\, a^{\mu}\big\}\big).\nonumber
\end{eqnarray}
The free dynamics of the field with
 integer spin $s$ within  BRST approach (see e.g. \cite{BKr}, \cite{BR}) is described by the first-stage reducible gauge theory with the gauge invariant action given on the configuration space $M^{(s)}_{cl}$ whose dimension grows with the growth of $"s"$, thus, including the basic field $\phi_{\mu(s)}$ with  auxiliary
fields $\phi_{1\mu(s-1)},...$ of lesser than $s$ ranks. All these
fields are incorporated into the vector $|\chi\rangle_s$ and the dynamics is
encoded  by the action
\begin{eqnarray}
\label{PhysStatetot} \mathcal{S}^m_{0|s}[\phi,\phi_1,...]=
\mathcal{S}^m_{0|s}[|\chi\rangle_s] = \int d\eta_0 {}_s\langle\chi|
KQ|\chi\rangle_s,
\end{eqnarray}
where $\eta_0$, $Q$ and $K$ be respectively a zero-mode ghost field, complete BRST operator  and an operator defining the
inner product. The action (\ref{PhysStatetot}) is invariant under
the reducible gauge transformations
\begin{eqnarray}
\label{gauge trasnform}
\delta|\chi\rangle_s =  Q|\Lambda^0\rangle_s , \ \ \delta |\Lambda^0\rangle_s = Q|\Lambda^1\rangle_s
,  \ \ \delta |\Lambda^1\rangle_s =0,
\end{eqnarray}
with  $|\Lambda^0\rangle_s$, $|\Lambda^1\rangle_s$ to be the
vectors of zero-level and first-level gauge  parameters of
abelian gauge transformations (\ref{gauge trasnform}). The
  BRST operator  $Q$ is
constructed on the base of the  constraints  $l_0,\, {l}_1,\, {l}{}^{+}_1=  - \imath a^{+\nu}  \partial_\nu ,\,
{l}{}_{11},\, {l}{}^{+}_{11} = \frac{1}{2}a^{+\nu}a^{+}_{\nu}$ with the
Grassmann-odd  ghost operators $\eta_0,\, \eta_1^+,\, \eta_1,\,
\eta_{11}^+,$ $\eta_{11}$,  $ {\cal{}P}_0$,  $ \mathcal{P}_{1}$,  $
\mathcal{P}^+_{1},$ $\mathcal{P}_{11},\, \mathcal{P}^+_{11},$ and also with two pairs of auxiliary Grassmann-even oscillators (with absence of $d,\, d^+$ for massless case). $d,\, d^+$,  $b,\, b^+$.
The ghost  and auxiliary operators   satisfy the non-zero respective  anticommuting and commuting  relations
\begin{equation}\label{ghanticomm}
  \{\eta_0, \mathcal{P}_0\}= \imath,\   \ \{\eta_1, \mathcal{P}_1^+\}=\{\eta^+_1, \mathcal{P}_1\}= \{\eta_{11},   \mathcal{P}_{11}^+\}=\{\eta_{11}^+,   \mathcal{P}_{11}\}=1; \ [d,\, d^+]=[b,\, b^+]=1.
\end{equation}
The BRST operator has the form
\begin{eqnarray}
&& {Q} =
\eta_0l_0+\eta_1^+\check{l}_1+\check{l}_1^{+}\eta_1+
\eta_{11}^+\widehat{L}_{11}+\widehat{L}_{11}^{+}\eta_{11} +
{\imath}\eta_1^+\eta_1{\cal{}P}_0,
\label{Qctotsym}
\end{eqnarray}
where
\begin{eqnarray}
\hspace{-0.5ex}&\hspace{-0.5ex}&\hspace{-0.5ex} \big( \check{l}_1,\, \check{l}_1^{+}  \big) =  \big( {l}_1 + m d,\, {l}_1^{+}+ m d^+  \big), \ \
\big(\widehat{L}_{11} ,\,\widehat{L}{}^+_{11}\big) =  \big(
\check{L}_{11}+\eta_{1} \mathcal{P}_{1} , \,
\check{L}{}^+_{11}+\mathcal{P}^+_{1}\eta^+_{1} \big).
\label{extconstsp2}
\end{eqnarray}
Here, the parameter $h = h(s)=-s - \frac{d-5-\theta_{m,0}}{2}$, $(\epsilon, gh_H) Q = (1, 1)$ and
\begin{eqnarray} \label{extconstsp21}
\check{L}_{11}={l}_{11}- \theta_{m,0}(1/2)(d)^2+(b^+b+h)b,\,\, \ \  \check{L}{}^{+}_{11}={l}^+_{11}- \theta_{m,0}(1/2)(d^+)^2 +b^+.
\end{eqnarray}
The algebra of the operators $l_0$ ,$l_1$, $l^{+}_1, L_{11}, L_{11}^+,
G_0$ presents a semidirect sum  of two subalgebras (isometries of $\mathbb{R}^{1,d-1}$ and $so(1,2)$):
\begin{equation}\label{subalgebr}
[l_0, l^{(+)}_1] = 0, \ [l_1,l_1^+]=l_0 -m^2 \quad \mathrm{and} \quad  [\check{L}_{11},  \check{L}_{11}^+] = G_0,\
[G_0, \check{L}_{11}^{+}] = 2\check{L}_{11}^+
\end{equation}
with their non-vanishing independent  cross-commutators  $[l_1,\check{L}_{11}^+]=-l_1^+$,  $[l_1,G_{0}]=l_1$, where
number particles operator $G_0$ as part of  the spin operator
${\sigma}$, is defined as follows
\begin{eqnarray}
\hspace{-0.5ex}&\hspace{-0.5ex}&\hspace{-0.5ex}  {\sigma}  =   G_0+ \eta_1^+\mathcal{P}_{1}
-\eta_1\mathcal{P}_{1}^+  + 2(\eta_{11}^+\mathcal{P}_{11} -\eta_{11}\mathcal{P}_{11}^+)
\label{extconstsp3}\\
\hspace{-0.5ex}&\hspace{-0.5ex}&\hspace{-0.5ex} G_0=g_0 + \theta_{m,0}d^+d +2b^+b+ \frac{1}{2}+ h. \nonumber
\end{eqnarray}
Spin operator  selects the vectors
with definite spin value $s$
\begin{eqnarray}
 \hspace{-0.5ex}&\hspace{-0.5ex}&\hspace{-0.5ex} {\sigma} (|\chi\rangle_s,\,
 |\Lambda^0\rangle_s,\, |\Lambda^1\rangle_s)  = (0,0,0),
\label{extconstsp}
\end{eqnarray}
where the standard distribution for  Grassmann parities and the ghost numbers $gh_H$ of the these vectors are $(0,0)$, $(1,-1),$ $(0,-2)$
respectively.

All the operators  above act  in a total Hilbert space  $\mathcal{H}_{tot}$ with the inner product of the vectors depending
on all oscillators  $(B;B^+)$ = $(a^{\mu},b,d; a^{\mu+},b^+,d^+)$ and ghosts
%$\langle\bullet| \bullet\rangle$
\begin{eqnarray}
&& \langle\psi |\chi \rangle = \int d^d x \langle0|  \chi^*\big(B;\eta_0, \eta_1, \mathcal{P}_1,\eta_{11}, \mathcal{P}_{11}\big)\psi\big(B^+;\eta_0,\eta^+_1, \mathcal{P}^+_1,\eta^+_{11}, \mathcal{P}^+_{11}\big)|0\rangle.
\label{scalarprod}
\end{eqnarray}

The operators $Q, {\sigma}$ are supercommuting and  Hermitian with
respect to the inner product (\ref{scalarprod}) including the
operator $K$ (see e.g., \cite{Reshetnyak_con}, \cite{BPT}, \cite{BR})  being equal to $1$ on Hilbert subspace not depending on auxiliary $b, b^+$ operators
  \begin{align}\label{geneq}
   & Q^2 =  \eta_{11}^+\eta_{11} \sigma ,\ && A^+K =  KA,\ \  A\in\{Q, \sigma\}; \\
     &    K=1\otimes \sum_{n=0}^{\infty}\frac{1}{n!}(b^+)n|0\rangle\langle 0|b^n C(n,h(s)),
&&  C(n,h(s))\equiv \prod_{i=0}^{n-1}(i+h(s))
   \label{geneq2}
  \end{align}
The BRST operator $Q$  is nilpotent on the subspace with zero
eigenvectors $\mathcal{H}^s_{tot}$ ($\mathcal{H}^s_{tot} \subset \mathcal{H}_{tot}$)  for the spin operator $\sigma$ (\ref{extconstsp}).

The field $ |\chi\rangle_s$, the zero $|\Lambda^0\rangle_s$ and  the
first $|\Lambda^1\rangle_s$ level gauge parameters
 labeled by the symbol $"s"$ as eigenvectors
of the spin condition in  (\ref{extconstsp})  has the  decomposition  with
ghost-independent vectors $|\Phi_{...}\rangle_{s-...}$, $|\Xi_{...}\rangle_{s-...}$
\begin{eqnarray}
\hspace{-1em}&\hspace{-1em}&\hspace{-1em} |\chi\rangle_s  =
|\Phi\rangle_s+\eta_1^+\Big(\mathcal{P}_1^+|\Phi_2\rangle_{s-2}+\mathcal{P}_{11}^+|\Phi_{21}\rangle_{s-3} +\eta_{11}^+\mathcal{P}_1^+\mathcal{P}_{11}^+|\Phi_{22}\rangle_{s-6}\Big) \label{spinctotsym} \\ \hspace{-1em}&\hspace{-1em}&\hspace{-1em} \phantom{ |\chi^0_c\rangle_s} +\eta_{11}^+\Big(\mathcal{P}_1^+|\Phi_{31}\rangle_{s-3}+\mathcal{P}_{11}^+|\Phi_{32}\rangle_{s-4}\Big)+  \eta_0\Big(\mathcal{P}_1^+|\Phi_1\rangle_{s-1}+\mathcal{P}_{11}^+|\Phi_{11}\rangle_{s-2}  \nonumber \\ \hspace{-1em}&\hspace{-1em}&\hspace{-1em} \phantom{ |\chi^0_c\rangle_s}  +  \mathcal{P}_1^+\mathcal{P}_{11}^+\Big[ \eta^+_{1} |\Phi_{12}\rangle_{s-4}+\eta^+_{11} |\Phi_{13}\rangle_{s-5}\Big]\Big),\nonumber \\
%%%%%%%%%%%%%%%%%%%%%%%%%%%%%%%%%%%%%%%%%%%
\hspace{-1em}&\hspace{-1em}&\hspace{-1em} |\Lambda^0\rangle_s  =  \mathcal{P}_1^+  |\Xi\rangle_{s-1}+\mathcal{P}_{11}^+|\Xi_{1}\rangle_{s-2} +\mathcal{P}_1^+\mathcal{P}_{11}^+\Big(\eta_1^+|\Xi_{11}\rangle_{s-4} \label{parctotsym}
\\
%%%%%%%%%%%%%%%%%%%%%%%%%%%%%%%%%%%%%%%%%%%
\hspace{-1em}&\hspace{-1em}&\hspace{-1em} \phantom{|\chi^1\rangle_s}
 + \eta_{11}^+|\Xi_{12}\rangle_{s-5}\Big) +  \eta_0\mathcal{P}_1^+\mathcal{P}_{11}^+|\Xi_{01}\rangle_{s-3} ,
  \nonumber\\
%%%%%%%%%%%%%%%%%%%%%%%%%%%%%%%%%%%%%%%%%%%
\hspace{-1em}&\hspace{-1em}&\hspace{-1em} {|\Lambda^1\rangle_s}  =
\mathcal{P}_1^+\mathcal{P}_{11}^+|\Xi^{1}\rangle_{s-3}. \label{gpar1}
\end{eqnarray}
Here for massive (for massless $d^{+}=0$) higher spin field
\begin{eqnarray}\label{Phiphi}
  |\Phi_{n}\rangle_{s-m} &=&  \sum_{l=0}^{[(s-m)/2]}\frac{(b^+)^l}{l!}\sum_{k=0}^{s-2l-m}\frac{(d^+)^k}{k!}|\phi_{n|l,k}(a^+)\rangle_{s-k-2l-m} \ \mathrm{for} \ |\phi_{0|0,0}(a^+)\rangle_{s}\equiv |\phi\rangle_s,\\
   \label{Xiphi} |\Xi_{j}\rangle_{s-1-m} &=& \sum_{l=0}^{[(s-1-m)/2]}\frac{(b^+)^l}{l!}\sum_{k=0}^{s-2l-1-m}\frac{(d^+)^k}{k!}|\Xi_{j|l,k}(a^+)\rangle_{s-k-2l-m-1}.
\end{eqnarray}

To formulate  BRST-BV action in minimal sector we  introduce configuration space $M^{(s)}_{\min}$  parameterized by fields $\Phi^A_{\min}=(A^i, C^{\alpha_0}, C^{\alpha_1})$ with all classical fields $A^i$, $i=1,...,n$, zero- and first-level  ghost fields $C^{\alpha_0}, C^{\alpha_1}$, $\alpha_0=1,...,m_0$, $\alpha_1=1,...,m_1$   (with using condensed DeWitt notations \cite{DeWitt}).  The zero- and first-level Abelian  gauge transformations with arbitrary functions $ \xi^{\alpha_0}, \xi^{\alpha_1}$ on $\mathbb{R}^{1,d-1}$, Noether identities , functional dependence of the generators $R^i_{0|\alpha_0}$ of zero-level gauge transformations are given for  Lagrangian formulation of  free higher-spin field $\phi_{\mu(s)}$ (with auxiliary fields entering  in $A^i$) as follows
\begin{eqnarray}\label{gibv}
   && \delta  A^i = R^i_{0|\alpha_0} \xi^{\alpha_0},\qquad \delta \xi^{\alpha_0} = Z^{\alpha_0}_{0|\alpha_1} \xi^{\alpha_1},\quad  \epsilon(A^i, \xi^{\alpha_0}, \xi^{\alpha_1})=(\epsilon_i, \epsilon_{\alpha_0}, \epsilon_{\alpha_1})=\vec{0} ;\\
   &&  \mathcal{S}^m_{0|s}[|\chi\rangle_s] \frac{\overleftarrow{\delta} }{\delta  A^i}R^i_{0|\alpha_0}=0, \qquad \ R^i_{0|\alpha_0}Z^{\alpha_0}_{0|\alpha_1}\big|_{ \big( \mathcal{S}^m_{0|s} \frac{\overleftarrow{\delta} }{\delta  A^i}=0\big)}=0.\label{gibv1}
\end{eqnarray}
Then,   we should  combine the field vector $ |\chi\rangle_s $  and vectors of ghost fields  $ |C^0\rangle_s, |C^1\rangle_s$ (obtained from the vectors of gauge parameters with help of some  Grassmann-odd constants $\mu_0, \mu_1$)  into  \emph{generalized field  vector}
\begin{equation}\label{gfv}
  |\chi_{\min}\rangle_s = |\chi\rangle_s + |C^0\rangle_s+ |C^1\rangle_s , \quad  |C^l\rangle_s\prod_{j=0}^l\mu_j \equiv  |\Lambda^l\rangle_s
\end{equation}
with all Grassmann-even terms in $|\chi_{\min}\rangle$.
Here the component tensors in the decomposition in power of ghost oscillators in the gauge parameters (\ref{parctotsym}), (\ref{gpar1}) one should change  on the respective
ghost tensor fields with the same rank but with shifted Grassmann parity and  Lagrangian ghost number $gh_L$:
\begin{eqnarray}\hspace{-1em}&\hspace{-1em}&\hspace{-1em} |C ^0\rangle_s  =  \mathcal{P}_1^+  |C^0_\Xi\rangle_{s-1}+\mathcal{P}_{11}^+|C^0_{\Xi{1}}\rangle_{s-2} +\mathcal{P}_1^+\mathcal{P}_{11}^+\Big(\eta_1^+|C^0_{\Xi{11}}\rangle_{s-4} \label{Cparctotsym}
\\
%%%%%%%%%%%%%%%%%%%%%%%%%%%%%%%%%%%%%%%%%%%
\hspace{-1em}&\hspace{-1em}&\hspace{-1em} \phantom{|\chi^1\rangle_s}
 + \eta_{11}^+|C^0_{\Xi{12}}\rangle_{s-5}\Big) +  \eta_0\mathcal{P}_1^+\mathcal{P}_{11}^+| C^0_{\Xi{01}}\rangle_{s-3} ,
  \nonumber\\
%%%%%%%%%%%%%%%%%%%%%%%%%%%%%%%%%%%%%%%%%%%
\hspace{-1em}&\hspace{-1em}&\hspace{-1em} {|C^1\rangle_s}  =
\mathcal{P}_1^+\mathcal{P}_{11}^+|C^1_{\Xi{1}}\rangle_{s-3} \label{Cpar1}
\end{eqnarray}
with the same  representation (\ref{Xiphi}) for  $|C^0_{\Xi{...}}\rangle$ and  $|C^1_{\Xi{1}}\rangle$ in terms of ghost fields $C^{\alpha_0}, C^{\alpha_1}$.
The fields, fields vectors,  ghost oscillators $\eta^{I}, \mathcal{P}_I$ satisfy to the Grassmann parity and ghost numbers distributions, for $(gh_H+gh_L)=gh_{\mathrm{tot}}$
\begin{eqnarray}\label{eghd}
 && \begin{array}{|c|cccccccc|} \hline
                   &   C^{\alpha_i}& |C^0_{\Xi{...}}\rangle  & |C^1_{\Xi{1}}\rangle  & |C^l\rangle & |\chi\rangle  &\eta^{I}& {\mathcal{P}}_{I} & \mu_l \\
\hline
                   \epsilon  &i+1 & 1 & 0 & 0 &0& 1 & -1 &1 \\
                   gh_H  & 0& 0 & 0 & -1-l & 0&1 & -1 & 0\\
                                      gh_L  &  i+1 & 1 & 2 & l+1 &0&0 & 0& -1 \\
                                                         gh_{\mathrm{tot}}  & l+1 & 1 & 2  & 0 &0& 1 & -1 &-1\\
                  \hline \end{array},\ \  i=0,1. \label{grassghtot}    \end{eqnarray}
Thus,  the generalized field  vector $ |\chi_{\min}\rangle_s$ has vanishing $(\epsilon, gh_{\mathrm{tot}})$ gradings and contains $2^4$
 ghost independent vectors $ |\Phi_{...}\rangle$, $|C^0_{\Xi{...}}\rangle, |C^1_{\Xi{1}}\rangle$ at independent ghost monomials
  \begin{equation}\label{decgfi}
\big\{ \eta_0^{n_0} (\eta_1^+)^{n_1}   (\eta_{11}^+)^{n_{11}} (\mathcal{P}_1^+)^{p_1}   (\mathcal{P}_{11}^+)^{p_{11}}\big\}, \ \mathrm{for} \ n_0, n_1, n_{11}, p_1, p_{11}=0,1.
 \end{equation}
with non-positive values of $gh_H$. Note, the inner product in $\mathcal{H}_{g}$ is  degenerate due to presence of Grassmann-odd field variables and we
 understand it as formal finite product according to  (\ref{scalarprod}), at least, for its values with vanishing $gh_L$.

BRST-BV action besides the field variables $\Phi^A_{\min}$  in the minimal sector depends on the same number of  antifields  $\Phi^*_{A|\min}= (A^*_i, C^*_{\alpha_0}, C^*_{\alpha_1} )$  organized in terms of respective vectors on the  space $\mathcal{H}_{g}$,
when considering instead of the  field vector $|\chi\rangle_s \in \mathcal{H}$ the \emph{generalized field-antifield vector} $|\chi_{g}\rangle_s  \in  \mathcal{H}_{g}$.
It contains in addition to  generalized field vector  the  \emph{generalized  antifield vector} $|\chi^*_{\min}\rangle_s$ with $2^4$
 ghost independent antifield vectors $ |\Phi^*_{...}\rangle$, $|C^{*0}_{\Xi{...}}\rangle, |C^{*1}_{\Xi{1}}\rangle$  at independent ghost monomials (\ref{decgfi}), but with positive
 values of $gh_H$
\begin{equation}\label{agfv}
 |\chi_{g}\rangle_s = |\chi_{\min}\rangle_s+|\chi^*_{\min}\rangle_s, \qquad  |\chi^*_{\min}\rangle_s = |\chi^*\rangle_s + |C^{*0}\rangle_s+ |C^{*1}\rangle_s
\end{equation}
The all  terms in $|\chi_{g}\rangle$ and in  $|\chi^*_{\min}\rangle$ are Grassmann-even.
Here
\begin{eqnarray}
\hspace{-1em}&\hspace{-1em}&\hspace{-1em} |\chi^*\rangle_s  =
\eta_0\Big\{|\Phi^*\rangle_s +\mathcal{P}_1^+\Big(\eta_1^+|\Phi^*_2\rangle_{s-2}+\eta_{11}^+|\Phi^*_{21}\rangle_{s-3} +\eta_1^+\eta_{11}^+\mathcal{P}_{11}^+|\Phi^*_{22}\rangle_{s-6}\Big) \label{aspinctotsym} \\
\hspace{-1em}&\hspace{-1em}&\hspace{-1em} \phantom{ |\chi^0_c\rangle_s} +\mathcal{P}_{11}^+\Big(\eta_1^+|\Phi^*_{31}\rangle_{s-3}+\eta_{11}^+|\Phi^*_{32}\rangle_{s-4}\Big)\Big\}+  \eta_1^+|\Phi^*_1\rangle_{s-1}+\eta_{11}^+|\Phi^*_{11}\rangle_{s-2}  \nonumber \\
\hspace{-1em}&\hspace{-1em}&\hspace{-1em} \phantom{ |\chi^0_c\rangle_s}  +  \eta_1^+\eta_{11}^+\Big[ \mathcal{P}^+_{1} |\Phi^*_{12}\rangle_{s-4}+\mathcal{P}^+_{11} |\Phi^*_{13}\rangle_{s-5}\Big],\nonumber \\
%%%%%%%%%%%%%%%%%%%%%%%%%%%%%%%%%%%%%%%%%%%
\hspace{-1em}&\hspace{-1em}&\hspace{-1em} |C^{*0}\rangle_s  =  \eta_0\Big\{\eta_1^+  |C^{*0}_{\Xi{1}}\rangle_{s-1}+\eta_{11}^+|C^{*0}_{\Xi{1}}\rangle_{s-2} +\eta_1^+\eta_{11}^+\Big(\mathcal{P}_1^+|C^{*0}_{\Xi{11}}\rangle_{s-4} \label{aparctotsym}
\\
%%%%%%%%%%%%%%%%%%%%%%%%%%%%%%%%%%%%%%%%%%%
\hspace{-1em}&\hspace{-1em}&\hspace{-1em} \phantom{|\chi^1\rangle_s}
 + \mathcal{P}_{11}^+|C^{*0}_{\Xi{12}}\rangle_{s-5}\Big) \Big\}+\eta_1^+\eta_{11}^+|C^{*0}_{\Xi{01}}\rangle_{s-3} ,
  \nonumber\\
%%%%%%%%%%%%%%%%%%%%%%%%%%%%%%%%%%%%%%%%%%%
\hspace{-1em}&\hspace{-1em}&\hspace{-1em} {|C^{*1}\rangle_s}  =
\eta_0\eta_1^+\eta_{11}^+|C^{*1}_{\Xi{1}}\rangle_{s-3}, \label{agpar1}
\end{eqnarray}
with the antifield vectors having the representation  in power of oscillators $a^+_{\mu}$, $d^+$, $b^+$ given by (\ref{Phiphi}), (\ref{Xiphi})
in terms of  antifield tensor fields with antifield $\phi_{\mu(s)}^*$ including in $|\Phi^*\rangle_s$ as coefficient for vanishing $d^+, b^+$.
The antifields   and composed from it  respective  antifield vectors obey to  the Grassmann parity and ghost numbers  gradings
 \begin{eqnarray}
 && \begin{array}{|c|cccccccc|} \hline
                   &A^*_i & C^*_{\alpha_l}&  |\Phi^*_{...}\rangle  & |C^{*0}_{\Xi{...}}\rangle  &   |C^{*1}_{\Xi{1}}\rangle &  |\chi^*\rangle & |C^{*l}\rangle &  |\chi^*_{\min}\rangle \\
                    \hline
                   \epsilon  &1&i& 1  & 0 & 1 & 0 & 0& 0 \\
                   gh_H  & 0 &0&0&  0 & 0 & 1& 2+l &  - \\
                                      gh_L  &-1&-l-2& -1 & -2 & -3 &  -1 & -2-l &   - \\
                                                         gh_{\mathrm{tot}}  &-1&-l-2&  -1 & -2 & -3   & 0 & 0&  0  \\
                  \hline \end{array},\ \  l=0,1  \label{grassghtotant}.    \end{eqnarray}
with undetermined values of $gh_{H}, gh_{L}$  for $ |\chi^*_{\min}\rangle$.  It provides due to vanishing of $(\epsilon, gh_{\mathrm{tot}})$ when calculated for the vectors $ |\chi_{\min}\rangle_s$, $ |\chi^*_{\min}\rangle_s$, the correctness of joint description of all field-antifield  variables $(\Phi^A_{\min}, \Phi^*_{A|\min})$  within one vector  $ |\chi_{g}\rangle_s$.

We see, that as it has already been  observed for the constrained BRST-BV approach the number of all  monomials composed from hamiltonian ghost  oscillators (\ref{decgfi}) and one of all ghost-independent vectors $ |\Phi^{(*)}_{...}\rangle$,  $ |C^{(*)0}_{\Xi{...}}\rangle$,  $ |C^{(*)1}_{\Xi{1}}\rangle$ coincide and equal to $2^5$ for our model.

Now,  BRST-BV minimal action $S^{(s)}_{\min}[\Phi_{\min}, \Phi^*_{\min}]$ in terms of field-antifield representation
\begin{equation}\label{Smin}
  S^{(s)}_{\min}[\Phi_{\min}, \Phi^*_{\min}]\ =\  \mathcal{S}^m_{0|s}[A] + A^*_i R^i_{0|\alpha_0} C^{\alpha_0}+ C^*_{\alpha_0} Z^{\alpha_0}_{0|\alpha_1} C^{\alpha_1}
\end{equation}
is determined  in  general vector form :
\begin{eqnarray}\label{Sgenfin1}
 &&  S^{(s)}_{\min}[\Phi_{\min}, \Phi^*_{\min}]  = {S}_{0|s}[|\chi_g\rangle_s] = \int d\eta_0 {}_s\langle\chi_g |
KQ|\chi_g\rangle_s,\\
 && \phantom{S^{(s)}_{\min}[\Phi_{\min}} =  \mathcal{S}_{0|s}[|\chi\rangle_s] + \int d \eta_0 \; \Big\{{}_{s}\langle \chi^*
|K Q |C^{0}\rangle_{s}   +  {}_s\langle C^{*0}|K Q |C^{1}\rangle_{s}
 + h.c.\Big\} \label{Sgenfin2}\end{eqnarray}
The functional ${S}_{0|s}[|\chi_g\rangle_s]$ has vanishing $\mathbb{Z}_2, \mathbb{Z}$- gradings: $(\epsilon, gh_H,  gh_L, gh_{\mathrm{tot}}){S}_{0|s}=\vec{0}$ and satisfies to the   master equation \cite{BV}
\begin{equation}\label{mestand0}
  \big({S}_{0|s}[|\chi_g\rangle_s], {S}_{0|s}[|\chi_g\rangle_s] \big)^{(s)}= 2{S}_{0|s}[|\chi_g\rangle_s] \frac{\overleftarrow{\delta}}{\delta\Phi^A_{\min}}\frac{\overrightarrow{\delta}}{\delta\Phi^*_{A|\min}}{S}_{0|s}[|\chi_g\rangle_s]=0
\end{equation}
 given in terms of odd Poisson bracket  (antibracket)  for any two differentiable  functionals $F,G$ determined  on the space $\mathcal{H}_g$ for given spin $s$ :
 \begin{equation}\label{oPb}
      \big(F[|\chi_g\rangle_s], G[|\chi_g\rangle_s] \big)^{(s)}\ =\  F[|\chi_g\rangle_s] \left( \frac{\overleftarrow{\delta}}{\delta\Phi^A_{\min}}\frac{\overrightarrow{\delta}}{\delta\Phi^*_{A|\min}}- \frac{\overleftarrow{\delta}}{\delta\Phi^*_{A|\min}}\frac{\overrightarrow{\delta}}{\delta\Phi^A_{\min}}\right) G[|\chi_g\rangle_s].
 \end{equation}
 The antibracket satisfies to the usual properties of bilinearity, generalized antisymmetry, Jacobi identity and Leibnitz rule for any $\alpha, \beta \in \mathbb{R}$ and $F,G,H$ given on $\mathcal{H}_g$ with definite  $\epsilon$,
\begin{eqnarray}\label{bila}
&& \big(\alpha F+\beta H, G \big)^{(s)} = \alpha\big( F, G \big)^{(s)}+\beta \big( H, G \big)^{(s)},\\
 && \big( F, G \big)^{(s)}=-(-1)^{(\epsilon(F)+1)(\epsilon(G)+1)}\big( G, F \big)^{(s)},\label{aatb} \\
 \label{Jacid}
 && (-1)^{(\epsilon(F)+1)(\epsilon(H)+1)}\big(\big( F, G \big)^{(s)},H \big)^{(s)} + \mathrm{cycle\  perm.}(F, G, H)=0,\\
 && \label{leibr} \big( FG, H\big)^{(s)} =F \big( G, H\big)^{(s)}+(-1)^{\epsilon(F)\epsilon(G)}G\big( F, H\big)^{(s)}.
\end{eqnarray}
As the reminiscent of gauge invariance of $\mathcal{S}_{0|s}[|\chi\rangle_s] $ the BRST-BV action  is invariant  with respect to the \emph{minimal Lagrangian BRST-like transformations} (with a Grassmann-odd constant parameter $\mu$, $(gh_H, gh_L)\mu =(0,-1)$) for the generalized field vector $|\chi_{\min} \rangle_s$
\begin{eqnarray}
  \delta_B|\chi_{\min} \rangle_{s}
& =  & \mu \frac{\overrightarrow{\delta}}{\delta \big({}_{s}\langle \chi^*_{\min}\big|K\big)}{S}_{0|s}[|\chi_g\rangle_s]  \ = \  \mu Q\big( | C^0 \rangle_{s}+ |C^1 ) \rangle_{s}\big),  \label{dxbrst0}  \\
 \delta_B |\chi^*_{\min} \rangle_{s}
& =  & 0 , \qquad \epsilon\left(\frac{\overrightarrow{\delta}}{\delta \big({}_{s}\langle\chi^{(*)}_{\min} \big|K\big)}\right) = 1,
\label{dxbrst1}
\end{eqnarray}
with constant antifields  (as well as for the duals $\langle \chi^{(*)}_{\min}|$)   or, equivalently, in terms of a \emph{BRST-like (Slavnov) generator} $\overrightarrow{s}_0$ and its dual $\overleftarrow{s}_0$:
 \begin{eqnarray}\label{brsnewgen}
  && \delta_B \left[ |\chi  \rangle_{s} , |C^0 \rangle_{s}, |C^1 \rangle_{s}\right]\ =\  \mu \overrightarrow{s}_0\left[|\chi (x) \rangle_{s} , |C^0 \rangle_{s}, |C^1 \rangle_{s}\right]  \ =\ \mu Q \left[ |C^0 \rangle_{s},\, |C^1 \rangle_{s},\, 0\right], \\
    && \delta_B \left[ {}_{s}\langle\chi|  , {}_{s}\langle C^0 |, {}_{s}\langle C^1 | \right]\ =\  \left[ {}_{s}\langle\chi|  ,\, {}_{s}\langle C^0 |,\, {}_{s}\langle C^1 | \right] \overleftarrow{s}_0 \mu\ =\ \left[  {}_{s}\langle C^0 |,\, {}_{s}\langle C^1 |,\, 0 \right] Q^+\mu . \label{brsnewgend}
\end{eqnarray}
Indeed,
\begin{eqnarray}\label{BRSTinv}
  \delta_B{S}_{0|s} &=&  \int d \eta_0 \left(\delta_B  {}_{s}\langle\chi_{\min}| \frac{\overrightarrow{\delta}S_{0|s}}{\delta {}_{s}\langle\chi_{\min} \big|}+ \frac{S_{0|s}\overleftarrow{\delta}}{\delta |\chi_{\min} \rangle_{s}\big|}\delta_B |\chi_{\min}\rangle_{s} \right) \\
   &=&  \mu \Big({}_{s}\langle \sum_i C^i | (Q^+)^2K  |\chi_g\rangle_{s}-{}_{s}\langle \chi_g| KQ^2  |\sum_iC^i\rangle_{s} \Big)=0.\nonumber
\end{eqnarray}
The variational derivatives   with respect to the vectors $ |\chi^{(*)}_{\min} \rangle_{s}$, and their duals in (\ref{dxbrst0}),  (\ref{dxbrst1}), (\ref{BRSTinv}), e.g.  for any quadratic (in the fields) functional with the kernel $E_{F}$
  \begin{equation}\label{represfunc}
  F  = \int d \eta_0 \mathcal{F}(\chi_g(\eta_0);\eta_0) = \int d \eta_0 {}_{s}\langle \chi_{g}
|K E_{F}| \chi_{g} \rangle_{s}
\end{equation}
are given in terms of  variational derivatives for a fixed  $\eta_0$ with a vanishing Grassmann parity of the density $\mathcal{F}$ ($\epsilon(\mathcal{F}) = \epsilon(E_{F})=\epsilon(F)+1$)  according to
\begin{eqnarray}
&\hspace{-1em}& \hspace{-1em}\left(\hspace{-0.3em}\frac{F \overleftarrow{\delta}}{\delta  |\chi^{(*)}_{\min} \rangle_s}; \frac{  \overrightarrow{\delta}F}{\delta {}_s\langle\chi^{(*)}_{\min} |}; \frac{F \overleftarrow{\delta}}{\delta  |C^{(*)j} \rangle_s}; \frac{  \overrightarrow{\delta}F}{\delta {}_s\langle C^{(*)j} |}\hspace{-0.3em}\right)   =   \left(\hspace{-0.2em}\frac{{\mathcal{F}}\overleftarrow{\delta}_{\eta_0}}{\delta  |\chi^{(*)}_{\min} \rangle_s};
\frac{  \overrightarrow{\delta}_{\eta_0}{\mathcal{F}}}{\delta {}_s\langle\chi^{(*)}_{\min} |}; \frac{\mathcal{F}\overleftarrow{\delta}_{\eta_0}}{\delta   |C^{(*)j} \rangle _s};
 \,\frac{  \overrightarrow{\delta}_{\eta_0}\mathcal{F}}{\delta {}_s\langle C^{(*)j}  |}\hspace{-0.2em}\right) \label{transf1}
\end{eqnarray}
(for $j=1,2$).  The  variational derivatives satisfy  the following normalization
conditions (for $\delta(\eta'_0-\eta_0) = \eta'_0-\eta_0$):
  \begin{align}
& \left( \frac{|B(\eta_0;x) \rangle_s  \overleftarrow{\delta}}{\delta  |B(\eta'_0;x') \rangle_s}; \frac{  \overrightarrow{\delta}{}_s\langle B (\eta_0;x)\big|}{\delta {}_s\langle B(\eta'_0;x')\big|}\right)  =  \delta(\eta'_0-\eta_0) \big(\delta(x'-x);  \delta(x'-x)\big), \  B \in \{{\chi}^{(*)}_{\min}, {C}^{(*)j}\}.\label{supus1}\end{align}

The construction of BRST-BV action    reflects the fact of BFV-BV duality \cite{BV-BFV}, \cite{GMR}  and realizes an example of AKSZ model \cite{AKSZ} when formulating BRST-BV minimal action with help of  hamiltonian  ghost-dependent Grassmann-odd  kernel (operator $Q$) which determines Lagrangian Grassmann-even gauge-invariant classical action and minimal BRST-BV action.

 We stress, first, that there are no gauge transformations for the  BRST-BV  functional ${S}_{0|s}[|\chi_g\rangle_s]$, because it encodes the gauge algebra functions itself. It makes  our approach to be  different from one  developed in \cite{BRST-BV3}.

 Second, as it was shown in \cite{BRcub}, \cite{BRcubmass}  after imposing the appropriate gauge conditions and eliminating the auxiliary fields from equations of motion for Lagrangian formulations, the
theory under consideration is reduced to Fronsdal form \cite{Fronsdal} for massless field of helicity $s$  in terms of
totally symmetric double traceless tensor field $\phi_{\mu(s)}$  and
traceless gauge parameter $\xi_{\mu(s-1)}$, as well as  to ungauge Singh-Hagen form \cite{SinghHagen} for massive field $\phi_{\mu(s)}$  of spin  $s$ with auxiliary  field $\phi_{3|\mu(s-3)}$. So that the respective BRST-BV minimal actions reduced to ones for irreducible gauge theory in  Fronsdal form and become non-gauge  for massive case.

Thus, we develop BRST-BV  approach to construct the minimal BRST-BV action for irreducible field with  integer spin $s$ in flat space-time.

Now we turn to the  interacting theory.

\section{Deformation procedure for interacting higher-spin fields}
\label{BRSTinter}

Here, we develop another prescription, firstly  for  the gauge theories initiated in \cite{BarnichHenneaux}, \cite{Henneaux1}, as compared to the  general (Noether-like) scheme  developed  in \cite{BRcub}, \cite{BRcubmass}  to find the cubic and higher
interaction vertices for the models with massless and massive  higher spin fields.

To  include the  interaction   we introduce $k$, $k\geq 3$ samples of Lagrangian formulations with vectors
$|\chi^{(j)}\rangle_{s_j}$, ghost fields
$|C^{(j)0}\rangle_{s_j}$, $|C^{(j)1}\rangle_{s_j}$ and respective antifield vectors combined in $k$-copies of  generalized field-antifield vectors $|\chi^{(j)}_{g}\rangle_{s_j}$ of the form (\ref{agfv}) with
corresponding vacuum vectors $|0\rangle^j$ and oscillators, where
$j=1,...,k$. It permits to  define the deformed BRST-BV action up to $p$-tic vertices, $p=3,4,...,e$  in  powers of deformation (also named as coupling) constant $g$ with preservation of its homogeneity in fields $|\chi^{(j)}_{g}\rangle_{s_j}$, starting from sum of $k$ copies of BRST-BV actions for free higher-spin fields and then from cubic, quartic and so on vertices:
\begin{eqnarray}\label{S[e]}
  && S^{(m)_k}_{[e]|(s)_k}[(\chi_g)_k] \ = \  \sum_{j=1}^{k} {S}^{m_j}_{0|s_j}[\chi^{(j)}_g]   + \sum_{f=1}^e g^f S^{(m)_k}_{f|(s)_k}[(\chi_g)_k];
  \end{eqnarray}
  where
\begin{eqnarray}\label{S[3]}
&&  S^{(m)_k}_{1|(s)_k}[(\chi_g)_k] =   \sum_{1\leq i_1<i_2<i_3\leq k}  \int \prod_{j=1}^3 d\eta^{(i_j)}_0  \Big( {}_{s_{i_j}}\langle \chi^{(i_j)}_g K^{(i_j)}
  \big|  V^{(3)}\rangle^{(m)_{(i)_3}}_{(s)_{(i)_3}}+h.c. \Big)  , \\
  \label{S[4]}
&&  S^{(m)_k}_{2|(s)_k}[(\chi_g)_k] =   \sum_{1\leq i_1<i_2<i_3<i_4\leq k}  \int \prod_{j=1}^4 d\eta^{(i_j)}_0  \Big( {}_{s_{i_j}}\langle \chi^{(i_j)}_g K^{(i_j)}
  \big|  V^{(4)}\rangle^{(m)_{(i)_4}}_{(s)_{(i)_4}}+h.c. \Big)  , \\
   && \ldots \ \ldots\ \ldots \ \ldots\ \ldots \ \ldots\ \ldots \ \ldots\ \ldots \ \ldots \ \ldots \ \ldots\ \ldots \ \ldots \nonumber\\
    \label{S[+]}
&&  S^{(m)_k}_{e|(s)_k}[(\chi_g)_k] =   \sum_{1\leq i_1<i_2<...<i_e\leq k}  \int \prod_{j=1}^e d\eta^{(i_j)}_0  \Big( {}_{s_{i_j}}\langle \chi^{(i_j)}_g K^{(i_j)}
  \big|  V^{(e)}\rangle^{(m)_{(i)_e}}_{(s)_{(i)_e}}+h.c. \Big)  ,
\end{eqnarray}
where we have used the notations $(\chi_g)_k= (\chi^{(1)}_g, \chi^{(2)}_g, ..., \chi^{(k)}_g)$ and the number of different terms in $p$-tic vertex is equal to $k!/((k-p)!p!)$.

In order to the interacting theory constructed from  initial BRST-BV actions ${S}^{m_j}_{0|s_j}$, $j=1,...,k$ would  preserve the number of physical  degrees of freedom $N_j$ determined by Lagrangian formulations for free higher-spin fields  with spin $s_j$, we demand  that the sum of all physical  degrees of freedom would be unchangeable, i.e. $\sum_{i}N_I = \mathrm{const}$. This property will be guaranteed if the deformed BRST-BV action $S^{(m)_k}_{[e]|(s)_k}$ will satisfy to
 the master equation  determined in terms of antibracket  given on the odd phase space $\Pi T^* M^{(s)_k}_{\min}$ (odd cotangent bundle) being locally presented as product of ones $\Pi T^* M^{(s_j)}_{\min}$ for each copy of higher-spin fields and parameterized by fields and antifields
  \begin{equation}\label{JPAP}
 \big(\Phi^{A_1}_{\min}, \Phi^{*}_{A_1|\min}, \Phi^{A_2}_{\min}, \Phi^{*}_{A_2|\min},\ldots, \Phi^{A_k}_{\min}, \Phi^{*}_{A_k|\min} \big) \equiv \big(\Phi^{Aa}_{\min}, \Phi^{*}_{Aa|\min}\big), \quad a=1,\ldots , k.
 \end{equation}
   The respective antibracket,  $\big(\bullet,  \bullet \big)^{(s)_k}$, is simply the sum of antibrackets  $\big(\bullet,  \bullet \big)^{(s_j)}$ for all copies. It has the form similar to (\ref{oPb}) with change: $\big(\Phi^{A}_{\min}, \Phi^{*}_{A|\min}\big)\to \big(\Phi^{Aa}_{\min}, \Phi^{*}_{Aa|\min}\big)$, thus obeying the same properties (\ref{bila})--(\ref{leibr}).

   The generating equation (master-equation) for deformed  BRST-BV action   to consistently find unknown  vertices $  \big|  V^{(3)}\rangle^{(m)_{(i)_3}}_{(s)_{(i)_3}}$, ... , $ \big|  V^{(e)}\rangle^{(m)_{(i)_e}}_{(s)_{(i)_e}}$
   \begin{equation}\label{mestand}
  \big(S^{(m)_k}_{[e]|(s)_k}[(\chi_g)_k] , S^{(m)_k}_{[e]|(s)_k}[(\chi_g)_k]  \big)^{(s)_k}= 2S^{(m)_k}_{[e]|(s)_k}[(\chi_g)_k] \frac{\overleftarrow{\delta}}{\delta\Phi^{Aa}_{\min}}\frac{\overrightarrow{\delta}}{\delta\Phi^*_{Aa|\min}}S^{(m)_k}_{[e]|(s)_k}[(\chi_g)_k] =0
\end{equation}
leads to the system of equations in powers of deformation constant $g$:
 \begin{eqnarray}
    &g^0:&   \big(S^{(m)_k}_{[0]|(s)_k}[(\chi_g)_k] , S^{(m)_k}_{[0]|(s)_k}[(\chi_g)_k]  \big)^{(s)_k} =  \sum_{j\leq k}\big(S^{m_j}_{0|s_j}[\chi^{(j)}_g] , S^{m_j}_{[j]|s_j}[\chi^{(j)}_g]  \big)^{(s_j)}\equiv 0  , \label{g0}  \\
   &g^1:&2 \big(S^{(m)_k}_{[0]|(s)_k}[(\chi_g)_k] , S^{(m)_k}_{1|(s)_k}[(\chi_g)_k]  \big)^{(s)_k} =0   , \label{g1} \\
    &g^2:& 2 \big(S^{(m)_k}_{[0]|(s)_k}[(\chi_g)_k] , S^{(m)_k}_{2|(s)_k}[(\chi_g)_k]  \big)^{(s)_k}+ \big(S^{(m)_k}_{1|(s)_k}[(\chi_g)_k] , S^{(m)_k}_{1|(s)_k}[(\chi_g)_k]  \big)^{(s)_k}   =0 , \label{g2} \\
&& \ldots \ \ldots \  \ldots \  \ldots \ \ldots \ \ldots \ \ldots \ \ldots \ \ldots \ \ldots \ \nonumber \\
    &g^{e}:& 2\sum_{j=0}^{\left[e/2\right]} \big(S^{(m)_k}_{j|(s)_k}[(\chi_g)_k] , S^{(m)_k}_{e-j|(s)_k}[(\chi_g)_k]  \big)^{(s)_k} \left(1-\frac{1}{2}\delta_{j,e-j}\right)  =0 . \label{ge}
 \end{eqnarray}
 The system is resolvable due to nilpotency of Grassmann-odd operator determined by BRST-BV action for free higher-spin fields with spins $s_1,...,s_k$
 \begin{eqnarray}\label{W}
   W &:=& \big(S^{(m)_k}_{[0]|(s)_k}[(\chi_g)_k] ,  \  \  \big)^{(s)_k},\quad  W^2= \big(S^{(m)_k}_{[0]|(s)_k}[(\chi_g)_k] , \big(S^{(m)_k}_{[0]|(s)_k}[(\chi_g)_k] , \ \  \big)^{(s)_k}   \big)^{(s)_k}\\
    &=&   \frac{1}{2}\sum_j \big(\big(S^{(m)_k}_{[0]|(s)_k}[(\chi_g)_k] , S^{(m)_k}_{[0]|(s)_k}[(\chi_g)_k] \big)^{(s)_k} ,\ \  \big)^{(s)_k} =0. \label{W2}
 \end{eqnarray}
To prove (\ref{W2}) we have used the relation (\ref{g0}), Jacobi identity (\ref{Jacid}) and antisymmetry property (\ref{aatb}) applied for antibracket $(\bullet,\bullet)^{(s)_k}$.  The equation (\ref{g1}) has the form, $W  S^{(m)_k}_{1|(s)_k}=0$.  Its non-trivial solution is determined  by non-zero $W$-cohomologies  in the space of  local functionals with vanishing $(\epsilon, gh_H, gh_L, gh_{\mathrm{tot}})$ gradings of the form (\ref{S[3]}), i.e.  by a
 $W$-closed functional $S^{(m)_k}_{1|(s)_k}$ having cubic dependence on whole field variables.    The necessary condition for the existence of solution  follows  by induction in $e$ for the quartic and higher deformations
 \begin{eqnarray}\label{ndeform}
   && W  \left(S^{(m)_k}_{1|(s)_k}[(\chi_g)_k] , S^{(m)_k}_{1|(s)_k}[(\chi_g)_k] \right)^{(s)_k}   = 2 \left(WS^{(m)_k}_{1|(s)_k}[(\chi_g)_k] , S^{(m)_k}_{1|(s)_k}[(\chi_g)_k]  \right)^{(s)_k} = 0,
 \end{eqnarray}
 The equations for quartic (\ref{g2}) and higher vertices (\ref{ge})  have more complicated  form  and its  solutions may have non-local (in space-time derivatives) representation \cite{T1}.

 In terms of ghost-dependent oscillator  kernels, e.g. the subsystem (\ref{g0}), (\ref{g1})  is presented
 \begin{eqnarray}
    &g^0:&   (Q^{tot})^2 \big|_{(\sigma |\chi_g\rangle=0)} =0, \qquad  Q^{tot} = \sum_{j}Q^{(j)} \label{g0Q}  \\
   &g^1:&  Q^{tot} \big|  V^{(3)}\rangle^{(m)_{(i)_3}}_{(s)_{(i)_3}} =0, \qquad \sigma^{(i_j)}\big|  V^{(3)}\rangle^{(m)_{(i)_3}}_{(s)_{(i)_3}} \ =\ 0, \label{g1Q} \end{eqnarray}
   for $j=1,2,3$;  $1\leq\hspace{-0.15em} i_1\hspace{-0.15em}<\hspace{-0.15em}i_2\hspace{-0.15em}<\hspace{-0.15em}i_3\hspace{-0.15em}\leq\hspace{-0.15em} k$.  These equations for $k\hspace{-0.1em}=\hspace{-0.1em}3$ copies of interacting fi\-elds co\-incide with ones obtained from BRST approach, when the cubic vertices for cubic de\-for\-ma\-tion in fi\-eld of the classical action and for linear in field deformed generator of gauge tra\-n\-sformations for the fields, also for the zero-level  gauge parameter were equal \cite{BRcub}, \cite{BRcubmass}.

   In this regard, we stress that if we weaken the homogeneity in generalized  vector $|\chi_g\rangle$ of the deformed BRST-BV action and will find it within the decomposition in field, antifield, ghost, antighost field vectors, then the same vertices as for BRST approach will appear in  $S^{(m)_k}_{1|(s)_k}[(\chi_g)_k]$ and a deformation procedure becomes more general (see, for this issue \cite{BL1}).

  A local on space-time coordinates dependence  in the vertices $\big|  V^{(3)}\rangle^{(m)_{(i)_3}}_{(s)_{(i)_3}}$ means
\begin{equation}\label{xdep}
\big|  V^{(3)}\rangle^{(m)_{(i)_3}}_{(s)_{(i)_3}} = \prod_{i=1}^3 \delta^{(d)}\big(x -  x_{i}\big) V^{(3)}{}^{(m)_{(i)_3}}_{(s)_{(i)_3}}
  \prod_{j=1}^3 \eta^{(i_j)}_0 |0\rangle , \ \  \  |0\rangle\equiv \otimes_{e=1}^k |0\rangle^{e}
\end{equation}
(for $(\epsilon, gh_{\mathrm{tot}})V^{(3)}{}^{(m)_{(i)_3}}_{(s)_{(i)_3}}= (0,0)$). We have the conservation law: $\sum_{j=1}^3p^{(i_j)}_\mu  = 0$,  for the momenta associated with all vertices.

Again as for BRST approach \cite{BRcub}, the cubic deformation of BRST-BV action  (in condensed notations),
\begin{eqnarray}\label{cubBVA}
   S^{(m)_k}_{[1]|(s)_k}\big(\Phi^{a}_{\min}, \Phi^{*}_{a|\min}\big)& =& \sum_j  S^{(s_j)}_{\min}[\Phi^{(j)}_{\min}, \Phi^{*(j)}_{\min}] +  g \Big(S^{(m)_k}_{1|(s)_k}\big(A^{a}\big) \\
    &+& \sum_{b,c,d}^k\big[A^{*}_{ib} R^{ib}_{1|\alpha_0d, jc}A^{jc}C^{\alpha_0d}+\frac{1}{2} C^{*}_{\gamma_0b}F^{\gamma_0b}_{{\alpha_0c\beta_0d}}C^{\alpha_0c}C^{\beta_0d}\big]+ \mathcal{O}(C^{\alpha_1b})\Big)\nonumber
\end{eqnarray}
because of  the cubic term with structural function  $F^{\gamma_0b}_{{\alpha_0c\beta_0d}}$   leads to the closure the algebra  of  deformed  gauge transformations:at linear approximation in $g$
\begin{equation}\label{closalg}
  R^{ia}_{1|\alpha_0b}(A)\frac{\overleftarrow{\delta}}{\delta A^{jc}}R^{j}_{0|\beta_0}- ((\alpha_0b)  \leftrightarrow (\beta_0c)) = -  R^i_{0|\gamma_0}F^{\gamma_0a}_{{\alpha_0b\beta_0c}},
\end{equation}
for $R^{ib}_{[1]|\alpha_0c}(A)= R^i_{0|\alpha_0}\delta^b_c+g R^{ib}_{1|\alpha_0c, ja}A^{ja}$.

The generating  equation (\ref{mestand}) together with system of equations (\ref{g0})--(\ref{ge})  determine deformation procedure for construction
 the  minimal  BRST-BV action (homogeneous in vectors $|\chi^{(a)}_g\rangle$) for interacting theory of irreducible  massive and massless fields with spins $s_1,...,s_k$.  For cubic vertices the system is reduced to (\ref{g0Q}),  (\ref{g1Q}),  which together with algebra of deformed gauge
transformations (\ref{closalg}) permit to find the cubic
vertices in terms of oscillator-like hamiltonian operators.

\section{Solution of  BRST-BV equations for Cubic Vertices} \label{BRSTsolgen}

Here we will construct the general solution for the cubic
vertices in following cases for interacting higher spin fields: with $k, k\geq 3$ massless; with  one massive and $(k-1)$ massless   of spins $s_1,
s_2, ..., s_k$  earlier developed within BRST approach for $k=3$  in \cite{BRcub}, \cite{BRcubmass}. The consideration is adapted for the finding  of Yang--Mills type interaction with gauge group $SU(n)$, $n\geq 2$.

\subsection{Cubic vertices for massless fields } \label{BRSTsolgen0}

We consider the equations  (\ref{g0Q}),  (\ref{g1Q}) for the cubic vertices $\big|  V^{(3)}\rangle^{(0)_{(i)_3}}_{(s)_{(i)_3}}\equiv $ $\big|  V^{(3)}\rangle_{(s)_{(i)_3}}$.  With use of the rule, $[i_{j+3} = i_j]$,  notations for the triple of numbers $i_1, i_2, i_3$ and
\begin{equation}\label{notp}
  \widehat{p}{}^{(i_j)}_{\mu}\ =\ {p}^{(i_{j+1})}_{\mu}-p^{(i_{j+2})}_{\mu}, \  \widehat{\mathcal{P}}{}^{(i_j)}_0\ =\  \mathcal{P}^{(i_{j+1})}_0- \mathcal{P}^{(i_{j+2})}_0, \ \ j=1,2,3.
\end{equation}
In \cite{BRcub}, \cite{Rcubmasless} it was shown that the $(Q^{tot}, (\sigma)_k)$-closed  solution for the equations (\ref{g1Q}) is presented by  product of $Q^{tot}$-closed forms constructed, in part, from the first and third  orders differential monomials  in powers of oscillators  suggested in \cite{BRST-BV3} as $Q_c^{tot}$-closed solution of the equation (\ref{g1Q}) but  for incomplete BRST operator $Q_c^{tot}$ ($Q_c^{tot}$ = $Q^{tot}|_{\eta^{(ii)(+)}_{11}=b^{(i)(+)}=0 }$ without imposing the trace operators (\ref{extconstsp2}), (\ref{extconstsp21})), which are now  written as
\begin{eqnarray}
    && L^{(i_j)} \ = \ \widehat{p}{}^{(i_j)}_{\mu})a^{(i_j)\mu+} - \imath\widehat{\mathcal{P}}^{(i_j)}_0
  \eta_1^{(i_j)+} ,  \label{LrL1} \\
  && Z^{(i)_3} \ = \  L^{(i_1i_2)+}_{11}L^{(i_3)} + L^{(i_2i_3)+}_{11}L^{(i_1)} + L^{(i_3i_1)+}_{11}L^{(i_2)}. \label{LrZ1}
\end{eqnarray}
Here we have used the definition  $p^{(i)}_{\mu} = -i\partial^{(i)}_{\mu}$ and
\begin{eqnarray}
L^{(i_j i_{j+1})+}_{11} \ = \ \textstyle a^{(i_j)\mu+}a^{(i_{j+1})+}_{\mu} -
\frac{1}{2}\mathcal{P}^{(i_j)+}_1\eta_1^{(i_{j+1})+} -
\frac{1}{2}\mathcal{P}^{(i_{j+1})+}_1\eta_1^{(i_j)+}.\label{Lrr+0}
 \end{eqnarray}
 The operators $ L^{(i_j)}, Z^{(i)_3}$ are not commuted with trace operators $\widehat{L}{}^{(i_j)}_{11}$, $j=1,2,3$. Therefore,  respective $Q^{tot}$-closed forms constructed from $ L^{(i_j)}$ are
 \begin{eqnarray}\label{LrZLL}
  &&  \mathcal{L}^{(i_j)}_{k_{i_j}} \ = \  \sum_{r=0}^{[k_i/2]} (-1)^{r}({L}^{(i_j)})^{r-2j}\big(\widehat{p}{}^{(i_j)}_{\mu} \widehat{p}^{(i_j)\mu}\big)^{r} \frac{k_i!}{r! 2^r(k_i-2r))!} \frac{(b^{(i_j)+})^j}{C(j,h^{(i_j)}(s_{i_j}))}   .
\end{eqnarray}
(the equivalent polynomial representation for BRST-closed operators  $ \mathcal{L}^{(i_j)}_{k_{i_j}}$ see \cite{BRcub}, \cite{BRcubmass}).

Also, the solution contains   new two-, four- , ..., $[s_{i_j}
/2]$ forms in powers of oscillators, corresponding to the trace
operators for $
j=1,2,3$
\begin{equation}\label{trform}
U^{(s_{i_j})}_{r_{i_j}}\big(\eta_{11}^{(i_j)+},\mathcal{P}_{11}^{(i_j)+} \big) \
: = \
(\widehat{L}{}^{(i_j)+}_{11})^{(j_i-2)}\big\{(\widehat{L}{}^{+(i_j)}_{11})^2
+ r_{i_j}(r_{i_j}-1)\mathcal{P}_{11}^{(i_j)+}\eta_{11}^{(i_j)+}\big\}.
\end{equation}
The  representation for $Q^{tot}$-closed modification of the operator $Z^{(i)_3}$ (corresponding to Yang--Mills type interaction) from the equation  $[Q^{tot}, \mathcal{Z}^{(i)_3}\}|0\rangle =0$ may be different. We choose one obtained in \cite{BRcub}, \cite{Rcubmasless}.

Resuming, we find the general  parity invariant  solution for the equations  (\ref{g1Q})
\begin{eqnarray}\label{genvertex}
 \big|  V^{(3)}\rangle_{(s)_{(i)_3}} &=& |{V}{}^{M(3)}\rangle_{(s)_{(i)_3}}  + \sum_{(r_{i_1},r_{i_2},r_{i_3}) >0}^{([s_{i_1}/2],[s_{i_2}/2],[s_{i_3}/2])} U^{(s_{i_1})}_{r_{i_1}}U^{(s_{i_2})}_{r_{i_2}}U^{(s_{i_3})}_{r_{i_3}}|{V}{}^{M(3)}\rangle_{(s)_{(i)_3}-2(r_{i})_3},
   \end{eqnarray}
where the vertex ${V}{}^{M(3)}_{(s)_{(i)_3}}$ is
given with modified forms $\mathcal{L}^{(i_j)}_{k_{i_j}}$, $\mathcal{Z}^{(i)_3}_j$ instead of $\big({Z}^{(i)_3}\big)^j$ (\ref{LrZ1})
\begin{eqnarray}\label{Vmets}
{V}{}^{M(3)}_{(s)_{(i)_3}-2(r_{i})_3} & = &   \sum_{p}\big(\mathcal{Z}^{(i)_3}\big)_{1/2\{(s_{(i)_3}-2r_{(i)_3}) - p\}}\prod_{j=1}^3 \mathcal{L}^{(i_j)}_{s_{i_j}-2r_{i_j}-1/2(s_{(i)_3}-2r_{(i)_3}- p)} ,  \\
\left(s_{(i)_3} ,r_{(i)_3}\right) &= & \big(\textstyle\sum_{j}s_{i_j} , \ \textstyle\sum_j r_{i_j}\big).
\end{eqnarray}
and is numerated by the natural parameter $p$ subject to the
equations
 \begin{eqnarray}
 s_{(i)_3}-2r_{(i)_3}-2s_{(i)_3 \min}\leq p\leq s_{(i)_3}-2r_{(i)_3}, \ \ \ p=s_{(i)_3}-2r_{(i)_3} - 2t, \ t\in \mathbb{N}_0.
\end{eqnarray}
The quantity above $\mathcal{Z}^{(i)_3}_j$ is determined, e.g.  for $j=1$ as follows
\begin{eqnarray}
  && \mathcal{Z}^{(i)_3}\prod_{j=1}^3\mathcal{L}^{(j)}_{k_j} = {Z}^{(i)_3}\prod_{j=1}^3\mathcal{L}^{(j)}_{k_j} - \sum_{i=1}^3k_i\frac{b^{(i)+}}{h^{(i)}}\big[\big[\widehat{L}^{(i)}_{11},{Z}^{(i)_3}\big\},{L}^{(i)} \big\}\prod_{j=1}^3\mathcal{L}^{(j)}_{k_j-\delta_{ij}} \label{Z1cal}\\
&& +\sum_{i\ne e}^3 k_ik_e\frac{b^{(i)+}b^{(e)+}}{h^{(i)}h^{(e)}}\big[\widehat{L}^{(e)}_{11},\big[\big[\widehat{L}^{(i)}_{11},{Z}^{(i)_3}\big\},{L}^{(i)} \big\}{L}^{(e)} \big\}\prod_{j=1}^3\mathcal{L}^{(j)}_{k_j-\delta_{ij}-\delta_{ej}} \nonumber\\
&& - \sum_{i\ne e\ne o}^3k_ik_ek_o\frac{b^{(i)+}b^{(e)+}b^{(o)+}}{h^{(i)}h^{(e)}h^{(o)}}\big[\widehat{L}^{(o)}_{11},\big[\widehat{L}^{(e)}_{11},\big[\big[\widehat{L}^{(i)}_{11},{Z}^{(i)_3}\big\},{L}^{(i)} \big\} {L}^{(e)} \big\}{L}^{(o)} \big\}\prod_{j=1}^3\mathcal{L}^{(j)}_{k_j-1}. \nonumber
\end{eqnarray}
For $j>1 $ the expressions for $\mathcal{Z}_j$ can be derived according to the  result \cite{Rcubmasless}.

 The general cubic vertex  for $k$ irreducible massless  higher-spin fields represents the $(3+1)$-parametric family to be enumerated by  parameters $r_{i_1}, r_{i_2}, r_{i_3}$ respecting for traces and parameter $p$. The representation for the cubic vertex will be then used for explicit tensor representation for interacting   massless field   with spins $s$ with $(k-1)$ scalars.

\subsection{Cubic vertices for two massless fields and one massive field} \label{BRSTsolgen1}

In this   case  in order to be differed from previous consideration  any cubic vertex should  include  one massive field with mass $m$ and spin $s_k\equiv s$ with rest two any massless fields from set of $(k-1)$ samples.  The solution for the  equation   (\ref{g1Q})  for the cubic vertex $\big|  V^{(3)}\rangle^{(0,0,m)}_{(s)_{(i)_3}}\equiv $ $\big|  V^{(3)}\rangle^m_{(s)_{(i)_3}}$ we will seek again as product of $(Q^{tot}, (\sigma)_k)$-closed  forms, starting from ones given within constrained BRST method \cite{BRST-BV3}  in so called {\emph{Minimal derivative scheme}} known also from light-cone approach \cite{Metsaev0512}.  In addition to the first order in oscillators   operators $ L^{(i_j)}$, $j=1,2,3$  (\ref{LrL1}) for $i_3 \equiv k$  there are three  second order in oscillators  operators
\begin{eqnarray}
  &\hspace{-0.5em}&\hspace{-0.5em}
L^{(i_1{}i_2)+}_{11}  = a^{(i_1)\mu+}a^{(i_2)+}_{\mu} +\frac{1}{2m^2} L^{(i_1)}L^{(i_2)} -
\frac{1}{2}\mathcal{P}^{(i_1)+}_1\eta_1^{(i_2)+} -
\frac{1}{2}\mathcal{P}^{(i_2)+}_1\eta_1^{(i_1)+}; \label{Lrr+1} \\
 &\hspace{-0.5em}&\hspace{-0.5em}
L^{(i_2 k)+}_{11} =  a^{(i_2)\mu+}a^{(k)+}_{\mu} -\frac{1}{2m^2} L^{(i_2)}L^{(k)}+\frac{1}{2m} L^{(i_2)}d^{(k)+}
-
\frac{1}{2}\mathcal{P}^{(i_2)+}_1\eta_1^{(k)+} -
\frac{1}{2}\mathcal{P}^{(k)+}_1\eta_1^{(i_2)+}\hspace{-0.3em};\label{Lrr+2}
\\
 &\hspace{-0.5em}&\hspace{-0.5em}
L^{(k i_1)+}_{11}  =  a^{(k)\mu+}a^{(i_1)+}_{\mu}  - \frac{1}{2m^2}L^{(k)} L^{(i_1)}-\frac{1}{2m}d^{(k)+} L^{(i_1)}
-
\frac{1}{2}\mathcal{P}^{(i_1)+}_1\eta_1^{(k)+} -
\frac{1}{2}\mathcal{P}^{(k)+}_1\eta_1^{(i_1)+}\hspace{-0.3em}.\label{Lrr+3}
 \end{eqnarray}
        These operators  {do not introduce the divergences into the vertices},  are Grassmann-even with vanishing ghost numbers $gh_H, gh_L$.

Note, first,  the operators (\ref{LrL1})
   $L^{(i_j)}$ for $j=1,2$ are not BRST $Q^{tot}_c$-closed
as compared to      $L^{(k)}$, $Q^{tot}_c L^{(k)}|0\rangle =0$.  Therefore, the  operator $Z$ (\ref{LrZ1}) in question  is not $Q^{tot}_c$-closed and hence not $Q^{tot}$-closed.
In turn, the operators $L^{(i_ji_{j+1})+}_{11}$, $\widehat{L}{}^{(i_j)+}_{11}$ are $Q^{tot}_c$-  BRST-closed, but not $Q^{tot}$-closed due to
\begin{eqnarray}&&
\widehat{L}{}^{(i_j)}_{11}L^{(i_ji_{j+1})+}_{11}|0\rangle = 0,\quad
\widehat{L}{}^{(i_j)}_{11}(L^{(i_ji_{j+1})+}_{11})^2|0\rangle  \ne 0,  \quad j=1,2,3
\label{LrL13}
 \end{eqnarray}
 The same properties of trace non-invariance are valid for $L^{(i_j)}$:
 \begin{eqnarray}
&&    \widehat{L}{}^{(i_j)}_{11}(L^{(i_j)})^2 |0\rangle =  \eta^{\mu\nu}\widehat{p}{}^{(i_j)}_{\mu} \widehat{p}^{(i_j)}_{\nu}|0\rangle = (\widehat{p}{}^{(i_j)})^2|0\rangle \ne 0, j=1,2,3.\label{Lr11Lr2}
 \end{eqnarray}

Then, again we have the respective $Q^{tot}$-closed  trace operators $U^{(s_{i_j})}_{r_{i_j}}$  (\ref{trform})  (massless for $j=1,2$ and massive $i_3=k$)

One can immediately check that the modified operators
\begin{eqnarray}
\label{LrZ10}
  &&  \mathcal{L}^{(3)}_{1} \ = \  {L}^{(3)} - [\widehat{L}{}^{(3)}_{11},{L}^{(3)}\}\frac{b^{(3)+}}{h^{(3)}}     ,   \\
\label{LrZ2m0}
  && \widetilde{ \mathcal{L}}{}^{(3)}_{2} \ = \  (\mathcal{L}^{(3)})^2 -i\widehat{\mathcal{P}}{}_0^{(3)}\eta^{(3)+}_{11}-{\widehat{l}}{}_0^{(3)}  \frac{b^{(3)+}}{h^{(3)}}     ,   \\
  \label{LrZ2m}
  &&  \widetilde{\mathcal{L}}{}^{(3)}_{2k} \ = \ (\widetilde{\mathcal{L}}^{(3)}_{2})^k, \qquad \widetilde{ \mathcal{L}}{}^{(3)}_{2k-1} \ = \ (\widetilde{\mathcal{L}}^{(3)}_{2})^{k-1}\mathcal{L}^{(3)}_{1}  ,
\end{eqnarray}
are invariant with respect to trace operators: $[\widehat{L}{}^{(3)}_{11},\, \mathcal{L}^{(3)}_{1}\}=0 $, and therefore are  $Q^{tot}$-closed.

As to the problem (\ref{LrL13}), we see
that  any power  of the
forms $L^{(i_ji_{j+1})+}_{11}$  are not $Q^{tot}$-closed.

One can check the $Q^{tot}$-closeness  for the mixed-symmetry modified  forms
\begin{eqnarray}\label{Lr12312}
  &&  \mathcal{L}^{(i_ji_{j+1})+}_{11|1} \ = \    {L}^{(i_ji_{j+1})+}_{11} -\sum_{i_0} {W}^{(i_0)}_{(i_ji_{j+1})|0}\frac{b^{(i_0)+}}{h^{(i_0)}} +\frac{1}{2}\Big(\sum_{i_0\ne j_0}[\widehat{L}{}^{(j_0)}_{11}, {W}^{(i_0)}_{(i_ji_{j+1})|0}\} \\
 && \quad \times \frac{b^{(i_0)+}}{h^{(i_0)}} \frac{b^{(j_0)+}}{h^{(j_0)}} +\sum_{i_0}[\widehat{L}{}^{(i_0)}_{11}, {W}^{(i_0)}_{(i_ji_{j+1})|0}\}\frac{(b^{(i_0)+})^2}{h^{(i_0)}(h^{(i_0)}+1)} \Big)  ,   \nonumber
 \end{eqnarray}
 %%%%%%%%%%%%%%%%%%%%%%%%%%%%%
 and for its any powers  $(\mathcal{L}^{(i_ji_{j+1})+}_{11|1})^k$.

In (\ref{Lr12312})  the indices $i_0, j_0$ are running two values:  $i_j, i_{j+1}$; for $j=1,2,3$ and fixed  $i_j=1,...,k$, also we have used the notations
\begin{eqnarray}
&& [\widehat{L}{}^{(i_0)}_{11}, L^{(i_ji_{j+1})+}_{11}\} \ \equiv \ W^{(i_0)}_{(i_ji_{j+1})|0}\,,  \label{defW}
 \end{eqnarray}
 By the construction the calligraphic operators are traceless, because of the last terms  in (\ref{Lr12312})  in front of the maximal power in $b^{(i_j)+}$, i.e.  $[\widehat{L}{}^{(j_0)}_{11}, {W}^{(i_0)}_{(i_ji_{j+1})|0}\}$
depend on only $a^{(i_j)}_{\mu}, d^{(k)}, \eta/^{(i_j)}_1$,  $\mathcal{P}_1^{(i_j)}$  (annihilation) oscillators and therefore the compensation procedure is finalized.

Both  $ \mathcal{L}^{(3)}_{1}$ and $\mathcal{L}^{(i_ji_{j+1})+}_{11|1}$ are non-commutative and  depend both on creation and annihilation operators in order theirs any powers to be $Q^{tot}$-closed quantities.

As the result, the {solution} for the  parity invariant vertex  looks
\begin{eqnarray}\label{genvertexm}
   \big|V^{(3)}\rangle^m_{(s)_{(i)_3}} &=&\big|{V}{}^{M(3)}\rangle^m_{(s)_{(i)_3}}  + \sum_{(r_{i_1},r_{i_2},r_{i_3}) >0}^{([s_{i_1}/2],[s_{i_2}/2],[s_{i_3}/2])} U^{(s_{i_1})}_{r_{i_1}}U^{(s_{i_2})}_{r_{i_2}}U^{(s_{i_3})}_{r_{i_3}}\big|{V}{}^{M(3)}\rangle^m_{(s)_{(i)_3}-2(r_{i})_3},
   \end{eqnarray}
where the vertex $\big|{V}{}^{M(3)}\rangle^m_{(s)_{(i)_3}-2(r_{i})_3}$ was
defined in the paper \cite{BRST-BV3} with account for (\ref{xdep})  but with modified forms $\mathcal{L}^{(3)}_{k}$,  (\ref{LrZ2m})  and $\big(\mathcal{L}^{(i_ji_{j+1})+}_{11|1}\big)^{\tau^{(i_j+2)}}$ instead of $\big({L}^{(i_ji+1)+}_{11}\big)^{\tau^{(i_j+2)}}$ (\ref{Lrr+1})--(\ref{Lrr+3})
\begin{eqnarray}\label{Vmetsm}
{V}{}^{M(3)|m}_{(s)_{(i)_3}-2(r_{i})_3} & = &   \textstyle\sum_{p}\mathcal{L}^{(3)}_{p}\prod_{j=1}^3 \big(\mathcal{L}^{(i_ji_{j+1})+}_{11|1}\big)^{\tau_{i+2}} , \ \end{eqnarray}
and is $(3+1)$-parametric  family to be  enumerated by the natural triple $(r_{i_j})_3$ respecting for the number of  traces and $p$ for  the minimal order of derivatives, also subject to the
relations
 \begin{eqnarray}
 && \tau_{i_j} =   \frac{1}{2}\big(s_{(i)_3}-2r_{(i)_3}-p\big)-s_{i_j}  ,\ j=1,2; \qquad   \tau_{k} =   \frac{1}{2}(s_{(i)_3}-2r_{(i)_3}+p)-s+2r_{3}  , \\
 && {\max}\big( 0, (s-2r_{3})-\sum_{j=1}^2(s_{i_j}-2r_{i_j})\big) \leq p\leq s-2r_{3}- \big|s_{i_1}-2r_{i_1}- (s_{i_2}-2r_{i_2})\big| , \\
 &&  0\leq  2r_{i_j} \leq 2 [s_{i_j}/2] , \qquad s_{(i)_3}-2r_{(i)_3}-p= 2t, \ t\in \mathbb{N}_0.
\end{eqnarray}
Note, first, that multiplicative-like  representation for the vertex (\ref{genvertexm}), (\ref{Vmetsm}) is differed from one suggested in \cite{BRcubmass}, but equivalent to it. Second, for vanishing $r_{({i})_3}$ remaining parameter  corresponds to one in constrained BRST formulation \cite{BRST-BV3}.

Emphasize, that including of  the constraints $\widehat{L}^{(i)}_{11}$ responsible
for the traces into BRST operator means that the usual condition
of vanishing double traces of the fields is fulfilled only on-shell
as the consequence of free equations of motion. Off-shell the
(double) traces of the fields do not vanish. At the same time the vanishing of double (single) traces of the fields (ghost fields) for interacting higher spin fields
  is modified as compared to the case of free dynamics however with preservation of irreducibility for  any interacting (basic) field.

\section{Massless HS fields with  helicities $(0, 0, s)$} \label{examples}

Here, we consider the cubic vertices  for the case of two  massless scalars and tensor fields with  helicities $(0, 0, s)$ for $k=3$ in ghost-independent and tensor forms.

First, we have according to (\ref{genvertex}), (\ref{Vmets}) the $r$-parameter family of vertices for $r=1,...,[s/2]$ with restoring  the dimensional coupling constants $t_r$ ($\dim t_r$=$s+d/2 -3-2r$, in metric units providing a dimensionless of the classical action) and with  $h^{(i)}(s_i)= -s\delta_{i3}-(d-6)/2$,
\begin{eqnarray}\label{genvertex1}
   {V}{}^{(3)|0}_{(0,0,s)} &=&\sum_{r \geq0}^{[s/2]}t_rU^{(s)}_{r}  \mathcal{L}^{(3)}_{s-2r} = \sum_{r\geq0}^{[s/2]}t_r U^{(s)}_{r}\sum_{i=0}^{[(s-2r)/2]} (-1)^{i}({L}^{(3)})^{s-2(r-i)}\times \\
   && \times (\hat{p}^{(3)})^{2i} \frac{(s-2r)!}{i! 2^i(s-2r-2i)!} \frac{(b^{(3)+})^i}{C(i,h^{(3)})} , \nonumber
   \end{eqnarray}
   also with following decomposition in powers of $\eta_1^{(3)+}$  for the operators
  \begin{eqnarray}
\label{genvertexaux1}
  & \hspace{-0.9em}& \hspace{-0.9em}  \mathcal{L}^{(3)}_{k} \ = \ \mathcal{L}^{(3)0}_{k} - \imath  \widehat{\mathcal{P}}^{(3)}_0\eta_1^{(3)+}\big(\mathcal{L}^{(3)}_{k-1}\big)^{\prime}\ \equiv \ \mathcal{L}^{(3)}_{k}|_{\eta_1^{(3)+}=0}\\
    & \hspace{-0.9em}& \hspace{-0.9em}  \ \qquad  - \imath  \widehat{\mathcal{P}}^{(3)}_0\eta_1^{(3)+}\sum_{i=0}^{[k/2]} (-1)^{i}\big(\widehat{p}{}^{(3)}_{\mu}a^{(i)\mu+}\big)^{k-1-2i} (\hat{p}^{(3)})^{2i} \frac{k!}{i! 2^i(k-1-2i)!} \frac{(b^{(3)+})^i}{C(i,h^{(3)})}.\nonumber
\end{eqnarray}
The field-antifield structure of interacting theory is determined  by the generalized vectors
 \begin{equation}\label{fafex}
   |\chi^{(j)}_g\rangle_0 = |\chi^{(j)}_{\min}\rangle_0+ |\chi^{*(j)}_{\min}\rangle_0 =  \big(\phi^{(j)}(x) + \eta^{(j)}_0 \phi^{*(j)}(x) \big)|0\rangle, \  \   |\chi^{(3)}_g\rangle_s =  |\chi^{(3)}_{\min}\rangle_s +  |\chi^{*(3)}_{\min}\rangle_s \end{equation}
(for $ j=1,2$)  with representations (\ref{spinctotsym}),  (\ref{gfv})--(\ref{Cpar1}) for the vectors corresponding to  classical  $A^{i3}$, zero- $C^{\alpha_03}$, first-level $C^{\alpha_13}$ ghost fields and
with  ones  (\ref{agfv})--(\ref{agpar1}) for the vectors for  respective antifields, without "massive" $d^{(3)(+)}$ oscillators in the component vectors (\ref{Phiphi}).

  The minimal BRST-BV action
 $S^{(0)_3}_{1|(s)_3}[(\chi_g)_3]$ (\ref{S[3]})  in the first order approximation in $g$
 \begin{eqnarray}\label{S[ex]}
&\hspace{-0.5em}&\hspace{-0.5em}  S^{(0)_3}_{[1]|(s)_3}[(\chi_g)_3] \ = \
\int  d^d x\sum_{i=1}^2  \phi^{(i)}\Box \phi^{(i)}+
 \int d\eta_0\Big[ {}_s\langle\chi^{(3)} |
K^{(3)}Q^{(3)}|\chi^{(3)}\rangle_s \\
 &\hspace{-0.5em}&\hspace{-0.5em} \quad  + \Big\{{}_{s}\langle \chi^{*(3)}
|K^{(3)} Q^{(3)} \overrightarrow{s}_0|\chi^{(3)}\rangle_s     +  {}_s\langle C^{*0{(3)}}|K^{(3)} \overrightarrow{s}_0 |C^{0{(3)}}\rangle_{s}
 + h.c.\Big\}  \Big]+g S^{(0)_3}_{1|(s)_3}[(\chi_g)_3], \nonumber \\
&\hspace{-0.5em}&\hspace{-0.5em} S^{(0)_3}_{1|(s)_3} =  \int \hspace{-0.2em} \prod_{i=1}^3 d\eta^{(i)}_0   \Big( \Big\{{}_{s}\langle \chi^{(3)} K
  \big| {}_{0}\langle \phi^{(2)}
  \big| {}_{0}\langle \phi^{(1)}\big|{V}{}^{(3)}\rangle^{0}_{s} + \hspace{-0.2em}\sum_{j=1}^2{}_{0}\langle \phi^{*(j)}
  \big|K \overrightarrow{s}_1\big| \phi^{(j)} \rangle_{0}
 \Big\} + h.c. \hspace{-0.15em}\Big)\hspace{-0.15em}, \label{S[ex]1}
 \end{eqnarray}
 where,  the first line  in (\ref{S[ex]})  corresponds  to classical action for free fields, the second line contains antifield terms with generator of initial  BRST-like transformations (\ref{brsnewgen}) for classical $|\chi^{(3)}\rangle_s$ and ghost $ |C^{{(3)0}}\rangle_{s}$ fields. In turn, the first term in  (\ref{S[ex]1}) means for cubic interacting part of classical action $S^{(0)_3}_{1|(s)_3}[(\chi)_3]$ and the second one for deformed  generator of BRST-like transformations  $\delta_{1|B}$ with generator $\overrightarrow{s}_1$ (with notation $\delta_{[1]|B} =\delta_{0|B}+\delta_{1|B}$):
  \begin{eqnarray}\label{brsnewgen1}
  && \delta_{1|B} |\phi^{(j)} \rangle_{0}\ =\   \mu \overrightarrow{s}_1|\phi^{(j)}  \rangle_{0}  \ =\ - g\mu  \int  \prod_{i=1}^2  d \eta^{(j+i)}_0  {}_s\langle C^{(3)0} K^{(3)}
  \big|  {}_{0}\langle \phi^{(j+1)}
  \big|{V}{}^{(3)}\rangle^{0}_{(0,0,s)} , \\
    &&  \delta_{1|B}  {}_{0}\langle\phi^{(j)}|  \ =\ {}_{0}\langle\phi^{(j)}|  \overleftarrow{s}_1 \mu\ =\ - g \int  \prod_{i=1}^2  d \eta^{(j+i)}_0{}^{0}_{(0,0,s)}\langle {V}{}^{(3)} K^{(3)} \big| \phi^{(j+1)} \rangle_0 \big|C^{(3)0}\rangle_s
  \mu , \label{brsnewgend1}
\end{eqnarray}
with untouched transformations for $|\chi^{(3)}  \rangle $: $\delta_{1|B}|\chi^{(3)}  \rangle =0$.
Note, the deformed gauge transformations for the fields $|\phi^{(j)} \rangle$ are obtained by substitution: $\delta_{1}|\phi^{(j)} \rangle = \delta_{1|B} |\phi^{(j)} \rangle_{0}|_{(\mu    {}_s\langle C^{(3)0}  \big| ={}_s\langle \Lambda^{(3)}  \big|)}$ in accordance with (\ref{gfv}). The algebra of deformed gauge transformations (\ref{closalg}) remains by Abelian, i.e. $F^{\gamma_0a}_{{\alpha_0b\beta_0c}}=0$ in the first approximation in $g$.

 The interacting classical action $S^{(0)_3}_{[1]|(0,0,s)}[\phi^{(j)}, \chi^{(3)}]$ in  (\ref{S[ex]})  depends  on basic $\mathbb{R}$-valued  fields $\phi^{(1)}$, $\phi^{(2)}$ $\phi^{(3)}_{\mu(s)}$ and on auxiliary ones  $\phi^{(3)}_{\mu(s-2l)|0,l}$. for  $l=0,...,[s/2]$,   according to (\ref{Phiphi}).

In the  ghost-independent form the  BRST-BV action $S^{(0)}_{0|s}[\chi^{(3)}_g] $ for  free field $|\chi^{(3)}\rangle_s$ is completely described by classical action (with omitting label "${}^{(3)}$" below in (\ref{Sungh0m})--(\ref{Scon00}))
\begin{eqnarray}
&&  \mathcal{S}^{0}_{0|s} [\chi^{(3)}] \ = \
 \mathcal{S}^0_{C|s}[\chi^{(3)}_c]  - \bigg\{\Big({}_{s}\langle \Phi\big| K \check{L}{}^{+}_{11} +{}_{s-2}\langle \Phi_2  \big|K-{}_{s-3}\langle \Phi_{31}  \big| K {l}{}_1 \label{Sungh0m}\end{eqnarray}
  \begin{eqnarray} && \phantom{\mathcal{S}_{s}  =}-{}_{s-4}\langle \Phi_{32}  \big|  K\check{L}{}_{11} \Big) \big|\Phi_{11}\rangle_{s-2}+ \Big(-{}_{s-2}\langle \Phi _2 \big| K \check{L}{}^{+}_{11} -{}_{s-6}\langle \Phi_{22}  \big|K\check{L}{}_{11} \nonumber \\
       &&  \phantom{\mathcal{S}_{s}  =}   +{}_{s-3}\langle \Phi_{31}  \big|  K    {l}{}^{+}_1-{}_{s-4}\langle \Phi_{32}  \big|  K \Big) \big|\Phi_{12}\rangle_{s-4} + \Big({}_{s-4}\langle \Phi_{32} \big| K {l}{}^{+}_{1}  + {}_{s-6}\langle \Phi_{22}  \big|K {l}{}_{1} \nonumber\\
       &&  \phantom{\mathcal{S}_{s}   =} - {}_{s-3}\langle \Phi_{21}  \big| K \check{L}{}^{+}_{11}+\frac{1}{2}{}_{s-5}\langle \Phi_{13}  \big|  K \Big) \big|\Phi_{13}\rangle_{s-5}+ {}_{s-3}\langle \Phi_{21}  \big| K\Big(l_0 \big|\Phi_{31}\rangle_{s-3}\nonumber\\
             &&  \phantom{\mathcal{S}_{s}  =} -  \check{L}{}_{11} \big|\Phi_{1}\rangle_{s-1} \Big) -  \frac{1}{2} \Big({}_{s-6}\langle \Phi_{22}  \big| K l_0 \big|\Phi_{22}\rangle_{s-6} - {}_{s-4}\langle \Phi_{32}  \big|K l_0 \big|\Phi_{32}\rangle_{s-4}  \Big) +h.c.\bigg\}, \nonumber
                \end{eqnarray}
       and by action of the  Slavnov generator on the components of field vector  $\big|\chi_{\min} \rangle_{s}$
\begin{eqnarray}
              &&\hspace{-0.7em} \overrightarrow{s}_0 \big|\Phi\rangle_{s}\ =\  {l}{}^{+}_1|C^0_{\Xi}\rangle_{s-1} + \check{L}{}^{+}_{11} |C^0_{\Xi1}\rangle_{s-2}, \label{0gtrind}\\
               &&\hspace{-0.7em}  \overrightarrow{s}_0 \big|\Phi_1\rangle_{s-1}\ = \  l_0|C^0_{\Xi}\rangle_{s-1} + \check{L}{}^{+}_{11} |C^0_{\Xi01}\rangle_{s-3},   \label{0gtrind111}\\
              &&  \hspace{-0.7em} \overrightarrow{s}_0 \big|\Phi_2\rangle_{s-2}\  =\ {l}{}_1 |C^0_{\Xi}\rangle_{s-1}+ \check{L}{}^{+}_{11}  |C^0_{\Xi 11}\rangle_{s-4}  - |C^0_{\Xi1}\rangle_{s-2}, \label{0gtrind112}\\ &&
               \overrightarrow{s}_0\big|\Phi_{21}\rangle_{s-3} \  =  \  {l}{}_1|C^0_{\Xi1}\rangle_{s-2}-{l}{}_1^{+}|C^0_{\Xi11}\rangle_{s-4}-||C^0_{\Xi01}\rangle_{s-3},  \label{01gtrind}\\
              &&\hspace{-0.7em} \overrightarrow{s}_0 \big|\Phi_{22}\rangle_{s-6} \ = \ - \check{L}{}_{11}|C^0_{\Xi11}\rangle_{s-4} + {l}{}_1|C^0_{\Xi12}\rangle_{s-5}   ,  \label{0gtrind122}\\
              && \hspace{-0.7em}\overrightarrow{s}_0 \big|\Phi_{31}\rangle_{s-3} \ = \  \check{L}{}_{11}|C^0_{\Xi}\rangle_{s-1} + \check{L}{}^{+}_{11}|C^0_{\Xi12}\rangle_{s-5} ,   \label{02gtrind}\\
              &&    \hspace{-0.7em} \overrightarrow{s}_0 \big|\Phi_{32}\rangle_{s-4}\  =\  \check{L}{}_{11} |C^0_{\Xi1}\rangle_{s-2}  - {l}{}^{+}_1  |C^0_{\Xi 12}\rangle_{s-5} +  |C^0_{\Xi11}\rangle_{s-4}  ,   \label{0gtrind222}\\ &&   \delta_0 \big|\Phi_{11}\rangle_{s-2} \ = \  l_0|C^0_{\Xi1}\rangle_{s-2} -  {l}{}^{+}_{1}|C^0_{\Xi01}\rangle_{s-3} , \label{0gtrind1},\\
              %%%%%%%%%%%%%%%%%%%%
              &&    \hspace{-0.7em}\overrightarrow{s}_0 \big|\Phi_{12}\rangle_{s-4}\  =\  l_0|C^0_{\Xi11}\rangle_{s-4}  -  {l}{}_1  |C^0_{\Xi01}\rangle_{s-3}  , \label{0gtrind1123}\\
               && \hspace{-0.7em} \overrightarrow{s}_0 \big|\Phi_{13}\rangle_{s-5} \ = \  l_0|\Xi_{12}\rangle_{s-5} -  \check{L}{}_{11}|\Xi_{01}\rangle_{s-3} , \label{0gtrind11},\\
              %%%%%%%%%%%%%%%%%%%%
              &\hspace{-0.7em}&\hspace{-0.9em}\overrightarrow{s}_0 \left(|C^0_{\Xi}\rangle, |C^0_{\Xi1}\rangle,|C^0_{\Xi11}\rangle, |C^0_{\Xi12}\rangle, |C^0_{\Xi01}\rangle\right)  =  \left( - l^{+}_{11} -b^+  ,\,{l}{}^{+}_1 ,\, {l}{}_1,\,  \check{L}{}_{11} ,  l_0\right) \big|C^1_{\Xi}\rangle_{s-3} \label{1gtrind}
                      .  \end{eqnarray}
      Here  the functional $\mathcal{S}^0_{C|s}$   is  the action  for triplet formulation for   fields of helicities  $s,s-2,...., 1(0)$ (following to \cite{franciasag}) or for the the  field of spin  $s$ with additional off-shell traceless constraint,  but with  $ b^{+}$-dependence  in the triplet $ |\chi_c\rangle_s$ =  $|\chi\rangle_s\big|_{(\eta_{11}^{+}, \mathcal{P}_{11}^{+})=0}$:     %%%%%%
    %%%%%%    %%%%%%%%%%%%%
    \begin{eqnarray}\label{Scon00}
  && \mathcal{S}^0_{C|s}[\chi_c]  =  \left({}_{s}\langle \Phi  \big|    {}_{s-2}\langle \Phi_2\big| {}_{s-1}\langle \Phi_1\big|   \right) K\left(\begin{array}{ccc}
  l_0 &   0 & -{l}{}^{+}_1  \\
 0 & -l^{}_0 &  {l}{}^{}_1    \\
        -{l}{}^{}_1 & {l}{}^{+}_1&  1  \end{array}\right)
       \left(  \begin{array}{l}\big|\Phi\rangle_s\\ \big|{\Phi_2}\rangle_{s-2} \\ \big|{\Phi_1}\rangle_{s-1}   \end{array} \right) .
                \end{eqnarray}

The interacting part of action $S^{(0)_3}_{1|(0,0,s)}$
 is also written in the ghost-independent form
  \begin{eqnarray}\label{S[n]1ind9}
  && S^{(0)_3}_{1|(0,0,s)}[\phi^{(1)},\phi^{(2)}, \chi^{(3)}] \ = \    -  \textstyle\prod_{j=1}^3 \delta^{(d)}\big(x-  x_{j}\big) \Big( \Big\{{}_{0}\langle \phi^{(2)}
  \big| {}_{0}\langle \phi^{(1)}\big|\Big({}_{s}\langle \Phi^{(3)} K^{(3)}
    \big|\\
    && \quad  \times\sum_{r \geq0}^{[s/2]}t_r (\check{L}{}^{(3)+}_{11})^r  \mathcal{L}^{(3)0}_{s-2j}
    + {}_{s-2}\langle \Phi^{(3)}_2 K^{(3)}
    \big|\sum_{r \geq0}^{[s/2]-1}t_r(r+1) (\check{L}{}^{(3)+}_{11})^r  \mathcal{L}^{(3)0}_{s-2(r+1)} \nonumber \\
    && \quad + {}_{s-4}\langle \Phi^{(3)}_{32} K^{(3)}
    \big|\sum_{r \geq0}^{[s/2]-2}t_r(r+1)(r+2) (\check{L}{}^{(3)+}_{11})^r  \mathcal{L}^{(3)0}_{s-2(r+2)}\nonumber \\
    && \quad - {}_{s-6}\langle \Phi^{(3)}_{22} K^{(3)}
    \big|\sum_{r \geq0}^{[s/2]-3}t_r(r+1)(r+2)(r+3) (\check{L}{}^{(3)+}_{11})^r  \mathcal{L}^{(3)0}_{s-2(r+3)}\Big\}|0\rangle +  h.c. \Big) \nonumber  ,
  \end{eqnarray}
  jointly with Slavnov  generator of BRST-like transformation
 \begin{eqnarray}
         &\hspace{-0.5em}&\hspace{-1.0em} \overrightarrow{s}_{[1]} \big| \phi^{(1)} \rangle_{0}  =  -
g \prod_{i=2}^3 \delta^{(d)}\big(x_{1} -  x_{i}\big){}_{0}\langle \phi^{(2)}\big| \Big\{{}_{s-1}\langle C^{0(3)}_{\Xi}
   K^{(3)}
    \big|\sum_{r\geq0}^{[s-1/2]}t_r (\check{L}{}^{(3)+}_{11})^r\big(\mathcal{L}^{(3)}_{s-1-2j}\big)^\prime\label{cubgtrex3gi} \end{eqnarray}
  \begin{eqnarray}
    &&\hspace{-0.5em} \quad  - {}_{s-5}\langle C^{0(3)}_{\Xi12}
   K^{(3)}
    \big|  \sum_{j \geq0}^{[s-5/2]}t_j(j+1)(j+2) (\check{L}{}^{(3)+}_{11})^j\big(\mathcal{L}^{(3)}_{s-5-2r}\big)^\prime\Big\}|0\rangle , \nonumber \\
    &\hspace{-0.5em}&\hspace{-1.0em}\overrightarrow{s}_{[1]} \big| \phi^{(2)} \rangle_{0}  = - \overrightarrow{s}_{[1]}\big| \phi^{(1)} \rangle_{0}\vert_{\textstyle \big(\phi^{(1)}(x_1)\to \phi^{(2)}(x_2)\big)}\label{cubgtrex2gi} .
\end{eqnarray}
Presenting these expressions in the oscillator forms, first, for cubic in fields action (\ref{S[n]1ind9}), second, for quadratic in fields Slavnov generators  (\ref{cubgtrex3gi}), (\ref{cubgtrex2gi})), then calculating the underlying inner products
we get   for the action  with accuracy up to overall factor $(-1)^ss!$
  \begin{eqnarray}\label{S[n]1ind1f}
  &&  \hspace{-0.5em} S^{(0)_3}_{1|(0,0,s)}=  -2 \int d^dx \bigg[\sum_{r\geq0}^{[s/2]}t_r\sum_{i=0}^{[s/2-r]}
     \sum_{l \geq0}^{r} \frac{C(l+i,h(s))}{ C(i,h(s))}\frac{r!}{l!(r-l)!} \frac{(s-2r)!}{i!}  \\
   && \hspace{-0.5em} \times    \bigg\{\sum_{u=0}^{s-2(r+i)} \frac{(-1)^u}{u!(s-2(r+i)-u)!}\sum_{q=0}^i \sum_{t=0}^{i-q} \frac{(-1)^t C_i^{q,t}}{2^{2(r-l)+i-t}} \Big[\partial_{\nu_0}...\partial_{\nu_u}\Big(\Box^{q}\partial_{\nu_{u+1}}...\partial_{\nu_{u+t}}\phi^{(1)}\Big)  \Big]  \nonumber\\
   && \hspace{-0.5em} \times \Big[\partial_{\nu_{u+t+1}}...\partial_{\nu_{s-2r-2i+t}}\Big(\partial^{\nu_{u+1}}...\partial^{\nu_{u+t}}\Box^{i-q-t}\phi^{(2)}\Big)\Big]
    \bigg\}
\phi^{(3)\nu({s-2(l+i)})}_{l+i,0}\prod_{p=1}^{j-l}\eta_{\nu_{s-2(r+i-p)-1}\nu_{s-2(r+i-p)}}  \nonumber\\
     && \hspace{-0.5em}\quad +  \sum_{r\geq0}^{[s/2]-1}t_r\sum_{i=0}^{[s/2-r-1]}
     \sum_{l \geq0}^{r} \frac{C(l+i,h(s))}{ C(i,h(s))}\frac{r!}{l!(r-l)!}   \frac{(r+1)!(s-2(r+1))!}{r!i! } \nonumber \\
   && \hspace{-0.5em}\times    \bigg\{\hspace{-0.2em}\sum_{u=0}^{s-2(r+1+i)}\hspace{-0.7em} \frac{(-1)^u}{u!(s-2(j+1+i)-u)!}\sum_{q=0}^i \sum_{t=0}^{i-q}  \frac{(-1)^t C_i^{q,t}}{2^{2(r-l)+i-t}} \Big[\partial_{\nu_0}...\partial_{\nu_u}\Big(\Box^{q}\partial_{\nu_{u+1}}...\partial_{\nu_{u+t}}\phi^{(1)}\Big)  \Big]  \nonumber\\
   && \hspace{-0.5em} \times \Big[\partial_{\nu_{u+t+1}}...\partial_{\nu_{s-2(j+1)-2i+t}}\Big(\partial^{\nu_{u+1}}...\partial^{\nu_{u+t}}\Box^{i-q-t}\phi^{(2)}\Big)\Big]
    \bigg\}
\phi^{(3)\nu({s-2-2l-2i})}_{2| l+i,0}  \nonumber\\
   &&\hspace{-0.5em} \quad \times\prod_{p=1}^{r-l}\eta_{\nu_{s-2(r+1+i-p)-1}\nu_{s-2(r+1+i-p)}} \nonumber
  \\
    &&\hspace{-0.5em}
     \quad +  \sum_{r \geq0}^{[s/2]-2}t_r\sum_{i=0}^{[s/2-r-2]}\sum_{k \geq0}^{r}
     \sum_{l \geq0}^{r} \frac{C(l+i,h(s))}{ C(i,h(s))}\frac{r!}{l!(r-l)!}  \frac{(r+2)!(s-2(r+2))!}{r!i! } \nonumber \\
   && \hspace{-0.5em}\times  \bigg\{\hspace{-0.3em}\sum_{u=0}^{s-2(r+2+i)}\hspace{-0.7em} \frac{(-1)^u}{u!(s-2(r+2+i)-u)!}\sum_{q=0}^i \sum_{t=0}^{i-q} \frac{ (-1)^t C_i^{q,t} }{2^{2(r-l)+i-t}} \Big[\partial_{\nu_0}...\partial_{\nu_u}\Big(\Box^{q}\partial_{\nu_{u+1}}...\partial_{\nu_{u+t}}\phi^{(1)}\Big)  \Big]  \nonumber\\
   && \hspace{-0.5em}\times \Big[\partial_{\nu_{u+t+1}}...\partial_{\nu_{s-2(j+2)-2i+t}}\Big(\partial^{\nu_{u+1}}...\partial^{\nu_{u+t}}\Box^{i-q-t}\phi^{(2)}\Big)\Big]
    \bigg\}
\phi^{(3)\nu({s-2(l+2)-2i})}_{32| l+i,0}  \nonumber\\
   &&\hspace{-0.5em} \quad \times\prod_{p=1}^{r-l}\eta_{\nu_{s-2(r+2+i-p)-1}\nu_{s-2(r+2+i-p)}}\nonumber \\
    && \hspace{-0.5em}\quad - \sum_{r\geq0}^{[s/2]-3}t_r\sum_{i=0}^{[s/2-r-3]}
     \sum_{l \geq0}^{r} \frac{C(l+i,h(s))}{ C(i,h(s))}\frac{r!}{l!(r-l)!}   \frac{(r+3)!(s-2(r+3))!}{r!i! }\nonumber \\
   && \hspace{-0.5em}\times  \bigg\{\hspace{-0.3em}\sum_{u=0}^{s-2(r+3+i)}\hspace{-0.7em} \frac{(-1)^u}{u!(s-2(r+3+i)-u)!}\sum_{q=0}^i \sum_{t=0}^{i-q} \frac{(-1)^t C_i^{q,t}}{2^{2(r-l)+i-t}} \Big[\partial_{\nu_0}...\partial_{\nu_u}\Big(\Box^{q}\partial_{\nu_{u+1}}...\partial_{\nu_{u+t}}\phi^{(1)}\Big)  \Big]  \nonumber
    \end{eqnarray}
   \begin{eqnarray}
   && \hspace{-0.5em}\times \Big[\partial_{\nu_{u+t+1}}...\partial_{\nu_{s-2(r+3)-2i+t}}\Big(\partial^{\nu_{u+1}}...\partial^{\nu_{u+t}}\Box^{i-q-t}\phi^{(2)}\Big)\Big]
    \bigg\}
 \phi^{(3)\nu({s-2(l+3+i)})}_{22| l+i,0}  \nonumber\\
   &&\hspace{-0.5em}  \quad \times\prod_{p=1}^{r-l}\eta_{\nu_{s-2(r+3+i-p)-1}\nu_{s-2(r+3+i-p)}}\bigg] \nonumber  ,
  \end{eqnarray}
(for the naturals $C^{k,l}_r\equiv \frac{r!}{k!l!(r-k-l)!}$ and  convention $\partial_{\nu_0}\equiv \prod_{i=1}^0 \partial_{\nu_i}  \equiv 1$), also for the  generators of BRST transformations
   \begin{eqnarray}
         &\hspace{-0.5em}&\hspace{-1.0em} \overrightarrow{s}_{[1]} \phi^{(1)}(x_1) =
          -g \int d^dx \bigg[\sum_{r \geq0}^{[(s-1)/2]}\hspace{-0.5em}t_r\hspace{-0.5em}\sum_{i=0}^{[(s-1)/2-r]}
     \sum_{l \geq0}^{r} \frac{C(l+i,h(s))}{ C(i,h(s))}\frac{r!}{l!(r-l)!} \frac{(s-1-2r)!}{i!}  \label{cubgtrex3gicomp11} \\
   &\hspace{-0.5em}&\hspace{-0.5em} \times    \bigg\{\sum_{u=0}^{s-1-2(r+i)} \frac{1}{u!(s-1-2(r+i)-u)!}\sum_{q=0}^i \sum_{t=0}^{i-q}  \frac{(-1)^t C_i^{q,t}}{2^{2(r-l)+i-t}}   {C_{\Xi}^{0(3)}}^{\nu({s-1-2(l+i)})}_{l+i,0} (x) \nonumber\\
   &\hspace{-0.5em}&\hspace{-1.0em} \times \Big[\partial_{\nu_{u+t+1}}...\partial_{\nu_{s-1-2(r+i)+t}}\Big(\partial^{\nu_{u+1}}...\partial^{\nu_{u+t}}\Box^{i-q-t}\phi^{(2)}\Big)\Big] \Big[\partial_{\nu_0}...\partial_{\nu_u}\Big(\Box^{q}\partial_{\nu_{u+1}}...\partial_{\nu_{u+t}}\Big)  \Big] \nonumber\\
   &\hspace{-0.5em}&\hspace{-1.0em} \quad \times\prod_{p=1}^{r-l}\eta_{\nu_{s-2(r+i-p)-2}\nu_{s-1-2(r+i-p)}}
    \bigg\}\delta^{(d)}\big(x -  x_{1}\big)
     \nonumber\\
  &\hspace{-0.5em}& \hspace{-1.0em}\quad  -
  \sum_{r \geq0}^{[(s-5)/2]}t_r\sum_{i=0}^{[(s-5)/2-r]}
     \sum_{l \geq0}^{r} \frac{C(l+i,h(s))}{ C(i,h(s))}   \frac{r!}{l!(r-l)!} \frac{(s-5-2r)!}{i!}  \nonumber \\
   &\hspace{-0.5em}& \hspace{-1.0em}\times    \bigg\{\sum_{u=0}^{s-5-2(r+i)} \frac{1}{u!(s-5-2(r+i)-u)!}\sum_{q=0}^i \sum_{t=0}^{i-q} \frac{(-1)^t C_i^{q,t} }{2^{2(r-l)+i-t}}  {C^{0(3)}_{\Xi12}}^{\nu({s-5-2(l+i)})}_{12|{}l+i,0} (x) \nonumber\\
   &\hspace{-0.5em}& \hspace{-1.0em}\times \Big[\partial_{\nu_{u+t+1}}...\partial_{\nu_{s-5-2(r+i)+t}}\Big(\partial^{\nu_{u+1}}...\partial^{\nu_{u+t}}\Box^{i-q-t}\phi^{(2)}\Big)\Big] \Big[\partial_{\nu_0}...\partial_{\nu_u}\Big(\Box^{q}\partial_{\nu_{u+1}}...\partial_{\nu_{u+t}}\Big)  \Big]\nonumber\\
   &\hspace{-0.5em}&\hspace{-1.0em} \quad \times\prod_{p=1}^{r-l}\eta_{\nu_{s-2(r+i-p)-6}\nu_{s-5-2(r+i-p)}}
    \bigg\}\delta^{(d)}\big(x -  x_{1}\big)
  \bigg]; \nonumber \\
   %%%%%%%%%%%%%%%%%%%%%%%%%%%%%%
    &\hspace{-0.5em}& \hspace{-1.0em}\overrightarrow{s}_{[1]} \phi^{(2)} (x_2) = -  \overrightarrow{s}_{[1]}\phi^{(1)} (x_1)|_{[ \phi^{(1)} (x_1) \to  \phi^{(2)} (x_2)]}. \label{cubgtrex2gicomp21}
\end{eqnarray}
Note, the vertex (\ref{S[n]1ind1f}) coincides with one  for one massive particle of spin $s$ interacting with two massless scalars in \cite{BRcubmass} for vanishing massive modes ($k=0$ in Eq. (127)).

 Let's simplify obtained solution for unconstrained interacting fields  by application of the gauge-fixing procedure (e.g. developed in \cite{BRcubmass}) of  auxiliary fields elimination for the vector $\chi^{(3)}$, because of the procedure  independence from the scalars $\phi^{(1)},\phi^{(2)}$. The classical action for massless tensor field is reduced to the form (\ref{Scon00}) for triplet of $b^{(3)+}$-independent fields  with BRST-transformations and subject to traceless constraints
 \begin{eqnarray}\label{traceres}
             && {l}_{11}\big|\Phi^{(3)}\rangle_s+  \big|\Phi^{(3)}_2\rangle_{s-2}=0,\quad  {l}_{11} \big|\Phi^{(3)}_k\rangle_{s-k}=0,\ k=1,2,\  {l}_{11}\big|C^{0(3)}_\Xi\rangle_{s-1}=0, \\
                     && \delta_{0|B} \left( \big|\Phi\rangle_{s} ,      \big|\Phi^{(3)}_1\rangle_{s-1},    \big|\Phi^{(3)}_2\rangle_{s-2}, \big|C^{0(3)}_\Xi\rangle_{s-1}\right) =\mu \left( {l}_1^+ ,l_0 , {l}_1, 0 \right) \big|C^{0(3)}_\Xi\rangle_{s-1}. \label{gtrd}
\end{eqnarray}

As the result, the interacting part of action (\ref{S[n]1ind1f}) will contain two terms with fields $\phi^{(3)\nu({s})}_{0,0}$ and $\phi^{(3)\nu({s-2})}_{2|0,0}$ without $b^{(3)+}$-generated  fields,
 so that, it is written as
     \begin{eqnarray}\label{S[n]1ind1fred}
  && S^{(0)_3}_{1|(0,0,s)} =  -2 \int d^dx \bigg[\sum_{r \geq0}^{1}t_r
     \frac{(s-2r)!}{2^{2r}}   \bigg\{\sum_{u=0}^{s-2r} \frac{(-1)^u}{u!(s-2r-u)!}
      \end{eqnarray}
      \begin{eqnarray}
        && \times\Big[\partial_{\nu_0}...\partial_{\nu_u}\phi^{(1)}  \Big] \Big[\partial_{\nu_{u+1}}...\partial_{\nu_{s-2r}}\phi^{(2)}\Big]
    \bigg\}
      \phi^{(3)\nu({s})}\prod_{p=1}^{r}\eta_{\nu_{s-2(r-p)-1}\nu_{s-2(r-p)}}  \nonumber\\
        && + t_0
        \bigg\{\sum_{u=0}^{s-2} \frac{(-1)^u(s-2)!}{u!(s-2-u)!} \Big(\partial_{\nu_0}...\partial_{\nu_u}\phi^{(1)}  \Big)  \Big(\partial_{\nu_{u+1}}...\partial_{\nu_{s-2}}\phi^{(2)}\Big)    \bigg\}
   \phi^{(3)\nu({s-2})}_{2| 0,0}\bigg], \nonumber
  \end{eqnarray}
  and generators of BRST-variations
   \begin{eqnarray} \label{cubgtrex3gicomp111}
         && \overrightarrow{s}_{[1]} \phi^{(1)}(x_1) =
          -g t_0\int d^dx
       \bigg\{\sum_{u=0}^{s-1} \frac{(s-1)!}{u!(s-1-u)!}   {C_{\Xi}^{0(3)}}^{\nu({s-1})}_{0,0} (x)\\
   &&\qquad  \times \Big(\partial_{\nu_{u+1}}...\partial_{\nu_{s-1}}\phi^{(2)}(x)\Big) \partial_{\nu_0}...\partial_{\nu_u}
    \bigg\}\delta^{(d)}\big(x -  x_{1}\big); \nonumber   \\
   %%%%%%%%%%%%%%%%%%%%%%%%%%%%%%
  && \overrightarrow{s}_{[1]} \phi^{(2)} (x_2) = -  \overrightarrow{s}_{[1]}\phi^{(1)} (x_1)|_{[ \phi^{(1)} (x_1) \to  \phi^{(2)} (x_2)]}. \label{cubgtrex2gicomp212}
\end{eqnarray}
    jointly  with the action (\ref{Scon00})  (also with ones for the scalars),  for free fields subject to the traceless constraints  (\ref{traceres})  may be served as interacting part of BRST-BV action for irreducible gauge theory  in the  triplet  formulation for the fields in question with double traceless initial  $ \phi^{(3)\nu({s})}$, traceless auxiliary $ \phi^{(3)\nu({s-k})}_k, k=1,2$ and ghost field $C^{0(3)\nu({s-1})}$.

Finally, expressing the triplet  in terms of only single field,  from the al\-geb\-raic equation of motion: $ \big|\Phi^{(3)}_1\rangle= {l}_1 \big|\Phi^{(3)}\rangle-{l}_1^+\big|\Phi^{(3)}_2\rangle$;  also from the constraint (\ref{traceres}):  ${l}{}^{(3)}_{11}\big|\Phi^{(3)}\rangle=-  \big|\Phi^{(3)}_2\rangle$ and substituting  in (\ref{Scon00}), (\ref{S[n]1ind1fred}), we get all components of BRST-BV action for  cubic approximation of interacting two scalars with double-traceless  field of spin $s$ in Fronsdal form with only one term in the vertex without trace due to the relation among the constants $t_1=2t_0$.

\section{Conclusion}

\label{Discus} %%%%%%%%%%%%%%%%%%%%%%%%%%%%%%%%%%%%%%%%%%%%%%%%

Summarizing, we have developed  BRST-BV approach with complete BRST operator $Q$ (\ref{Qctotsym})  as the augmentation of usual BRST approach for construction of
  the BRST-BV action in the minimal sector of field-antifield variables for the first-stage reducible gauge theory which describes free Lagrangian dynamics of totally-symmetric tensor field with integer spin  in $d$-dimensional Minkowski space-time.   To this aim we included all classical (initial and auxiliary), ghost fields and theirs antifields into unique Grassmann-even  generalized field-antiield vector $|\chi_g\rangle_s$ (\ref{agfv})  (with vanishing total ghost number) in generalized  space $\mathcal{H}_g$  (with degenerate inner product due to presence of Grassmann-odd field variables) as the component tensors in the decomposition in powers of ghost (creation), auxiliary $b^+, d^+$ and initial $a^+_{\mu}$  oscillators. This property reflects the equality of  the numbers of all  monomials composed from hamiltonian-like ghost  oscillators (\ref{decgfi}) and one of all ghost-independent vectors $ |\Phi^{(*)}_{...}\rangle$,  $ |C^{(*)0}_{\Xi{...}}\rangle$,  $ |C^{(*)1}_{\Xi{1}}\rangle$ in (\ref{spinctotsym}), (\ref{Cparctotsym}), (\ref{Cpar1}), (\ref{aspinctotsym}), (\ref{aparctotsym}), (\ref{agpar1})
            coincide and equal to $2^5$ for the model with number of all constraints $[o_i]=5$  determining $Q$. As the result the minimal BRST-BV action  with unrestricted fields and antifields encodes all structural functions of  Abelian gauge algebra of reducible gauge transformations and  has the same form as the classical BRST action (\ref{PhysStatetot}) if  we change the field $|\chi\rangle_s$ on $|\chi_g\rangle_s$ with the same BRST operator, which includes all differential and holonomic (related to trace)   constraints selecting irreducible representation (particle)  with higher integer spin $s$. The construction is adapted both for massless and massive irreducible fields. The  BRST-BV action coincides with the action constructed according BV method, thus satisfying to the master equation (\ref{mestand0}) in terms of component antibracket, and is invariant   with respect to BRST-like transformations (\ref{brsnewgen}), (\ref{brsnewgend}).  The concept of construction of BRST-BV action reveals the concept of BV-BFV duality \cite{BV-BFV}, \cite{GMR}  and realizes the example of AKSZ model \cite{AKSZ}.

            The additive deformation procedure for construction the minimal BRST-BV action  $\hspace{-0.1em}S^{(m)_k}_{[e]|\hspace{-0.1em}(s)_k}$  (\ref{S[e]}) for interacting $k$ copies of  fields with integer spins $s_1,...,s_k$  to be homogeneous in powers of generalized vectors $|\chi^{(i)}_g\rangle_{s_i}$ $i=1,...,k$  is developed as power series in deformation constant $g$, starting from sum of free   BRST-BV actions for each  copy, then in cubic (at $g^1$) approximation in  $|\chi^{(i)}_g\rangle_{s_i}$ with unknown cubic vertices, quartic   (at $g^2$) approximation and etc.  The requirement to preserve the number of physical degrees of freedom for the interacting theory is guaranteed by fulfillment the generating (master)-equation  (\ref{mestand})  for deformed BRST-BV action in total field-antifield space, whose regular in $g$ solution should  be  found from the  equivalent system of equations (\ref{g0})--(\ref{ge}) in powers of the constant $g$. It is shown, that the resolvability  of this system is based on the nilpotent Grassmann-odd operator $W$ related to the nilpotent total BRST operator $Q^{tot}$ (\ref{g0Q}). In terms of  BRST $Q^{tot}$ and spin operator  $(\sigma)_k$  the equation for cubic vertices for $S^{(m)_k}_{[1]|(s)_k}[(\chi_g)_k] $ is rewritten in equivalent oscillator form (\ref{g1Q}) (obtained earlier within BRST approach \cite{BRcub}, \cite{BRcubmass}) with unknown vertices $ \big|  V^{(3)}\rangle^{(m)_{(i)_3}}_{(s)_{(i)_3}}$. The general solution for the cubic  vertex, in order to have non-trivial interaction between the fields,  should be  $(Q^{tot}, (\sigma)_k)$-closed.

The  generic covariant local solution for the cubic
vertices is  constructed within BRST-BV approach for two cases of interacting higher spin fields: with $k, k\geq 3$ massless higher spins;  with  one massive and any two from $(k-1)$ massless   of spins $s_1, s_2, ..., s_k$  earlier developed within BRST approach \cite{BRcubmass} for $k=3$. For the first case,  the  parity invariant  cubic vertex $ \big|  V^{(3)}\rangle_{(s)_{(i)_3}}$ (with its $k!/[(k-3)!]3!$ its independent number) is given by the expressions (\ref{genvertex}),
(\ref{Vmets}) constructed from  BRST-closed differential forms  (\ref{LrZLL}), (\ref{Z1cal})  and the new forms (\ref{trform}) related to
the trace operator constraints.   The vertex has a
non-polynomial structure  and  presents  $(3+1)$-parameteric family to
be  enumerated by the naturals  $(r_{i_1},r_{i_2},r_{i_3})$  respecting for
the orders of traces incoming into the vertex, and $p$ enumerating
the order of derivatives in it.
In the second case for one massive and rest massless fields  the  parity invariant  cubic vertex $ \big|  V^{(3)}\rangle^m_{(s)_{(i)_3}}$  (\ref{genvertexm}), (\ref{Vmetsm})
 has multiplicative structure, composed from the non-commutative BRST-closed operators (\ref{LrZ10})--(\ref{LrZ2m}),
(\ref{Lr12312}), and trace-related forms (\ref{trform}). Again, the vertex  presents  $(3+1)$-parametric family with the same interpretation.

We have applied   the previous case to find new cubic (linear in $g$) deformation of minimal  BRST-BV action $ S^{(0)_3}_{1|(s)_3}[(\chi_g)_3]$ (\ref{S[ex]}) in ghost-independent and tensor forms
 for the triple of massless scalar  $\phi^{(1)}(x)$, $\phi^{(2)}(x)$ fields and massless tensor $\phi^{(3)}_{\mu(s)}(x)$  of integer helicity $s$.
 The initial and deformed parts of the BRST-BV action contain inputs into free  classical action $ \mathcal{S}^{0}_{0|s} [\chi^{(3)}] $ for massless tensor  (\ref{Sungh0m}) and for its cubic deformation, determined by the vertex $ {V}{}^{(3)|0}_{(0,0,s)}$ (\ref{genvertex1}) and  given by $S^{(0)_3}_{1|(0,0,s)}$  (\ref{S[n]1ind9}),
 In its turn, the action of initial Slavnov generator $\overrightarrow{s}_0$ (and its dual $\overleftarrow{s}_0$), which determine antifield-dependent part of BRST-BV action is  trivial for the massless scalars and for basic and auxiliary ghost-independent field and ghost vectors are given in (\ref{0gtrind})--(\ref{1gtrind}).
And vice versa the deformed   part of the Slavnov generator $\overrightarrow{s}_1$ (related to the deformed gauge transformations) for the latter is trivial, but arises    for the scalars (\ref{cubgtrex3gi}), (\ref{cubgtrex2gi}) and depend on two ghost-independent zero-level ghost field vectors. The tensor form
of deformed classical action and the action of the operator  $\overrightarrow{s}_1$  are given by (\ref{S[n]1ind1f}) and (\ref{cubgtrex3gicomp11}), (\ref{cubgtrex2gicomp21}). The structure of interacting theory permits one to apply partial  gauge-fixing procedure on the language of Lagrangian dynamics (independent from the scalars) and partial resolution of the equations of motion   to eliminate,  $b^+$-dependence from all quantities,   to remove all ghost fields with except for unique zero-level ghost field, ${C_{\Xi}^{0(3)}}^{\nu({s-1})}_{0,0} (x)$, also to remove all auxiliary fields with
 except for triplet of   fields: double-traceless $\phi^{(3)}_{\mu(s)}(x)$  and traceless $\phi^{(3)}_{k|\mu(s-k)}(x)$, $k=1,2$. For the latter triplet formulation the tensor representations for deformed part of BRST-BV action (\ref{S[n]1ind1fred}), (\ref{cubgtrex3gicomp111}), (\ref{cubgtrex2gicomp212}) permits to get the BRST-BV minimal action in terms of single double-traceless field, two scalars and traceless ghost field.

   We stress, that suggested BRST-BV  approach  with complete BRST operator due to the condition of homogeneity of the BRST-BV  action dependence in generalized  field-antifield vector
  should be evaluated as the less general method to find deformations of classical action. gauge transformations and gauge algebra as compared to the possibilities of  BRST approach
\cite{BRcub}, \cite{BRcubmass}. However without this condition both approaches become by equivalent.

From the obtained solutions it follows the possibilities  to
construct a  BRST-BV action in cubic approximation  for  interacting first-stage reducible
Lagrangian formulations for irreducible $(k-l)$ massless and $l$ massive fields, $l\geq 2$.

There are many directions for development and  application   of
the suggested procedure, such as finding quartic (vertices) approximation for BRST-BV action in whole field-antifield variables  for
irreducible massless integer higher spin fields on flat
backgrounds; for massive  integer higher spin fields, for higher
spin fields with mixed symmetry of indices, for higher spin
supersymmetric fields, where the vertices should include any powers
of traces.  One should also note the problems of
constructing the fourth and higher vertices and related
problems of locality (see the discussions in \cite{T1}, \cite{DT}, \cite{DGKV}, \cite{Vasil}, \cite{Didenko1}), where the BRST-BV procedure may  be useful.
The construction and  {quantum} loop calculations with the
BRST-BV quantum action for the models with derived cubic vertices can
be realized following to \cite{2010.15741}  within BRST-BV approach with complete BRST operator.

\vspace{-1ex}

\paragraph{Acknowledgements}
The author is grateful to I.L. Buchbinder,  S.A.  Fedoruk, Yu.M. Zinoviev, P.M. Lavrov,  to or\-ga\-nizers and participants of the VII International Conference “Models in Quantum Field The\-ory” in Saint Petersburg for  useful discussions and warm hospitality and also to referee for careful reading and many comments. The work was par\-ti\-ally sup\-por\-ted by the Ministry of
Education of Russian Federation, project No  QZOY-2023-0003.

\end{document}